\documentclass[11pt,a4paper]{article}
\usepackage{jheppub}
\usepackage{bm}
\usepackage{feynmf}
\usepackage{comment}
\usepackage{mathrsfs}
\usepackage{mathtools}
\usepackage[dvipsnames]{xcolor}
\usepackage{physics}
\usepackage{hyperref}
\usepackage{graphicx}
\usepackage{orcidlink}
\usepackage{listings}
\usepackage[utf8]{inputenc}
\usepackage{parskip}
\usepackage{soul}
\usepackage{amsfonts}
\usepackage{amssymb}
\usepackage{tensor}
\usepackage{float}
\usepackage{simpler-wick}
\usepackage{cancel}
\usepackage{amsthm}
\usepackage{tcolorbox}
\usepackage{enumitem}
\usepackage{siunitx}
\tcbuselibrary{most}

\usepackage{tikzducks}
\usepackage{marginfix}

\newcommand{\Mpld}{M_{\text{Pl},d}}
\newcommand{\MplD}{M_{\text{Pl},D}}
\newcommand{\Lsp}{\Lambda_\text{sp}}
\newcommand{\Nsp}{N_\text{sp}}
\newcommand{\DD}{\mathrm{D}}
\newcommand{\bphi}{\boldsymbol{\phi}}
\newcommand{\bPhi}{\boldsymbol{\Phi}}
\newcommand{\brho}{\boldsymbol{\rho}}

\theoremstyle{definition}
\newcommand{\thisconjecturename}{}
\newtheorem*{genericconjecture}{\sffamily \thisconjecturename}
\newenvironment{conjecture}[1]{
    \renewcommand{\thisconjecturename}{#1}%
    \begin{genericconjecture}}{
    \end{genericconjecture}
}

\tcolorboxenvironment{genericconjecture}{
    boxrule=0pt,
    boxsep=0pt,
    colback={White!90!Green},
    enhanced jigsaw,
    borderline west={2pt}{0pt}{Green},
    before skip=10pt,
    after skip=10pt,
    sharp corners,
    breakable
}

\tcolorboxenvironment{genericgreybox}{
    boxrule=0pt,
    boxsep=0pt,
    colback={White!90!Gray},
    enhanced jigsaw,
    borderline west={2pt}{0pt}{Gray},
    before skip=10pt,
    after skip=10pt,
    sharp corners,
    breakable
}

\title{\Huge From Frame Covariance to the Swampland Distance Conjecture}

\author[a]{Sotirios Karamitsos \orcidlink{0000-0002-1773-249X}}
\author[b,c]{and Benjamin Muntz \orcidlink{0000-0002-0183-8783}}
\date{~}

\affiliation[a]{Institute of Physics, University of Tartu, W. Ostwaldi 1, 50411 Tartu, Estonia}

\affiliation[b]{School of Physics and Astronomy, University of Nottingham, University Park, Nottingham NG7 2RD, United Kingdom}

\affiliation[c]{Nottingham Centre of Gravity, University of Nottingham, University Park, Nottingham NG7 2RD, United Kingdom}

\emailAdd{sotirios.karamitsos@ut.ee}
\emailAdd{benjamin.muntz@nottingham.ac.uk}

\abstract{
Field space geometry plays a central role within the Swampland Programme, most notably in the various Distance Conjectures. However, for gravitational EFTs, this geometry is not uniquely defined: one can cast the action in many synonymous descriptions related by Weyl transformations, in which the field space metric transforms non-trivially across conformal frames. This raises a crucial question of how we are meant to think of the field space metric in view of employing the Swampland Conjectures. In this work we resolve this ambiguity by developing a fully frame-covariant framework for studying gravitational EFTs. We show that all conformal frames arise as distinct foliations of a singular higher-dimensional auxiliary geometry. Applying ADM formalism to the augmented field space, it is clear how Weyl- and unit transformations can be understood from a geometric point of view. Using this framework, we revisit the Species Scale Distance Conjecture and Sharpened Distance Conjecture, and show how the bounds derive from universal properties of gravitational EFTs under Weyl transformations. This strongly suggests that aspects of these conjectures apply to a much broader class of scalar-tensor theories and are consequences of frame covariance, rather than constraints imposed by quantum gravity. 
}

\begin{document}
\maketitle
\flushbottom

\section{Introduction}
Our universe and phenomena within it --- at least the way we observe them at low energies --- are accurately modelled and tested against a vast landscape of effective field theories (EFTs). Even so, there is a prevailing expectation that any such EFT, if it is to describe a physically realisable world, must ultimately be compatible with quantum gravity in the sense of admitting a consistent UV completion. Over the past decade, much effort has been devoted to uncovering what the word `compatibility' might entail. This has led to the formulation of a growing collection of proposed constraints, collectively known as Swampland Conjectures. The central aspiration of the Swampland Programme is that such constraints may serve as a filter. That is, to organise low-energy EFTs into those that can be consistently embedded into quantum gravity, and those that cannot; the latter are said to lie in the ``Swampland''. By applying these conjectures to cosmology, particle physics, and beyond, one hopes to infer phenomenological restrictions that might otherwise appear inconspicuous from low-energy considerations alone. 

Naturally, such a quest is delicate. Many facets and features of quantum gravity itself remain obscure, and it is not always clear-cut which properties of known quantum-gravitational constructions are fundamental and which are artefacts of particular frameworks (cf.~\cite{Buoninfante:2024yth} where this is debated). The web of Swampland Conjectures has grown impressive in scope --- slowly but surely painting a bigger picture --- yet their origins and domain of validity can sometimes appear murky. This invites a more cautious responsibility to interrogate such conjectures rather than merely deploy them. Where do they truly stem from? Are there hidden underlying assumptions in play? And, perhaps most importantly, do they genuinely reflect deep principles of \emph{quantum} gravity, or could they be consequences of more pedestrian aspects of our formulations of gravitational EFTs? This paper takes the cautious stance.

One cornerstone of the web of proposed constraints concerns the family of Distance Conjectures, which generically assert that large excursions in field space are accompanied by discernible physical effects --- typically the emergence of towers of exponentially light states that signal the breakdown of the original EFT. These statements are often formulated in terms of the geodesic distance traversed by fields (during, say, cosmological evolution), and therefore implicitly rely on the existence and well-definedness of a metric on the field manifold. While the precise details vary between versions of the conjectures, the common thread is the idea that \emph{distance traversed in field space} is a meaningful, physically consequential quantity. Understanding and rigidly defining a notion of distance is therefore essential to assess and employ the claims of the Swampland Programme.

That being said, the geometry of field space in \emph{gravitational} EFTs is subtler than one might first expect. Gravitational theories typically admit multiple equivalent descriptions related by Weyl transformations, and these conformal frames generally assign \emph{different} metrics to the underlying field space. The mapping between metrics is non-trivial and a na\"ive pullback will not, in general, preserve things like geodesics and distances. This simple observation alone should motivate the need for a frame-covariant understanding and formulation of the Distance Conjectures. A further conceptual dilemma arises from the fact that the Distance Conjectures are usually phrased in `$d$-dimensional Planck units' --- even though physical statements should not depend on the choice of units. The reliance on Planck units, put simply, risks encumbering the intended physical significance. In fact, frame covariance is closely connected to the question of unit (in)dependence, as we elaborate on later in this paper. Although the Swampland literature inherently deals with gravitational EFTs and routinely makes use of Weyl transformations to go between conformal frames,\footnote{See \cite{Kehagias:2019akr, Lust:2019zwm} for related uses of Weyl transformations within the Swampland Programme and in the context of the AdS Distance Conjecture, with the purpose of defining generalised notions of distance that includes the spacetime metric.} the implications of frame covariance for the Distance Conjectures have remained largely unattended to. The aim of this paper is therefore to fill this gap: by adopting a frame-covariant point of view, we will lay out a novel formalism to grasp the field space geometry of gravitational EFTs, and through this new lens, revisit some of the Distance Conjectures.

The structure of this paper is as follows: Sections~\ref{sec:swampland} and~\ref{sec:fieldspacegeometry} serve as overviews of the two topics we plan to bridge. In Section~\ref{sec:swampland}, we provide an introduction to the Swampland Programme and some of the Distance Conjectures, intended for the reader mainly familiar with scalar-tensor theories. Likewise, in Section~\ref{sec:fieldspacegeometry}, we introduce the concept of frame covariance with a particular focus on Weyl- and unit transformations in the context of scalar-tensor theories, meant for the more Swampland-familiar audience. Section~\ref{sec:augmentedfieldspace} is the heart of this paper: we formulate the idea of the frame-augmented field space and provide a geometric picture of everything discussed prior, including conformal frames, Weyl- and unit transformations, as well as field space geodesics and geodesic distance. Our framework is, in principle, applicable far beyond the Swampland Programme. Notwithstanding, we aim our attention at and revisit the Species Scale Distance Conjecture and Sharpened Distance Conjecture in Section~\ref{sec:distanceconjecture}, using toroidal compactifications as a pedagogical example. We hope that this will set the stage for further studies and invite healthy discussions between the different communities. Appendix~\ref{sec:appendix} contains further examples to demonstrate one of our main results in Eq.~\eqref{eq:speciesscaleclaim}.

\section{A Pedestrian Guide to the Swampland Distance Conjecture}
\label{sec:swampland}
In this section we will provide a pedestrian overview of the Swampland Programme and some of its associated conjectures. Our goal is to paint an elementary picture for those who are not already versed in the Swampland literature, including those readers arriving from the arena of frame covariance or scalar-tensor theories. To that end, we will go over the basic concepts and the established terminology and notation of the Swampland Programme.

The grand goal in fundamental physics is to identify the low-energy EFT which accurately describes physics as we experience it. To put it briefly, the Swampland is the domain of low-energy EFTs coupled to gravity that do \emph{not} admit a UV completion in a theory of quantum gravity. Charting the Swampland appears at first glance to be an impossible task, as we have no universal consensus on what quantum gravity ought to fully entail. A major development was Vafa's observation that, in a measure-theoretic sense, `most' EFTs coupled to gravity will lie in the Swampland \cite{Vafa:2005ui}. This implied that there must be non-trivial, and likely quite restrictive, constraints on the set of allowed EFTs. If these constraints can be articulated, they can in turn be used to make phenomenological predictions.\footnote{In this sense, the Swampland Programme shares many ideological features with for example the Positivity Bounds and Asymptotic Safety programmes. The overlap between these different approaches \cite{Basile:2021krr, Basile:2025zjc} --- that is, the EFTs that are ruled out by each of the respective criteria --- has also colloquially been dubbed the ``Absolute Swampland'' \cite{Eichhorn:2024rkc}.} 

Over the past decade, a plethora of constraints have been put forward, dubbed Swampland Conjectures (until rigorously proven). While they differ in scope, they are similar in practice: if an EFT violates a Swampland Conjecture, it is considered to be in the Swampland.\footnote{This is not to say that Swampland Conjectures are absolute; after all, they are still conjectures, possibly relying on some unspecified underlying assumptions that can be evaded. In the case of an EFT violating a Swampland Conjecture, the EFT could either be in the Swampland or, alternatively, the EFT could provide a valid counterexample. In some instances, conjectures have also been retrospectively rephrased or ``refined'' to account for such outcomes \cite{Ooguri:2018wrx, Blumenhagen:2019vgj}. What one ought to conclude is ultimately a discussion beyond the scope of our paper, but we nevertheless find it important to remark on this scientific dilemma.}  For further details about how Swampland Conjectures are conceptualised, put to the test, as well as how they relate to one another, we refer the reader to general reviews such as \cite{Palti:2019pca, vanBeest:2021lhn, Grana:2021zvf, Lehnert:2025izp}.

In this work, we will focus on a subset of so-called Distance Conjectures, which derive from the original Swampland Distance Conjecture put forward by Ooguri and Vafa in \cite{Ooguri:2006in}. Now, at the time of writing there are several kinds of Distance Conjectures, and introducing each of them is excessive compared to what this paper will address. For our purposes, it is sufficient to explain that they all have one theme in common: they concern the behaviour of gravitational EFTs when one traverses the space of their vacuum solutions, and the parametric behaviour will be a function of the \emph{distance} travelled. In a string theory context, this space is referred to as the \emph{moduli space}, parameterised by the vevs of massless scalar fields (moduli) present in the EFT. To make things more concrete, consider the following $d$-dimensional EFT:
\begin{equation}
\label{sectEFT}
    S\supset \int \dd^dx\ \sqrt{-g}\left[ \frac{\Mpld^{d-2}}{2}R[g] - \frac{1}{2}g^{\mu\nu}\mathsf{G}_{ij}(\bphi)\partial_\mu\phi^i\partial_\nu\phi^j\right]\,.
\end{equation}
Here, the index $i=1,\dots,N$ labels the fields and hence $N$ is the dimension of the moduli space, which we denote $\mathcal{M}$. The kinetic function $\mathsf{G}_{ij}(\bphi)$ is then identified as the metric on $\mathcal{M}$. In string theory (and in particular when dealing with compactifications), the moduli space essentially encodes the geometry of extra spatial dimensions. This is because, if we start with a gravitational theory on a $D$-dimensional manifold $\mathfrak{M}_d\times \mathfrak{X}_{D-d}$ with $\mathfrak{X}_{D-d}$ compact, we can obtain a lower-dimensional EFT on $\mathfrak{M}_d$ by integrating out $\mathfrak{X}_{D-d}$ (a common example in string theory is $D=10$, $d=4$, and $\mathfrak{X}_6$ a six-dimensional Calabi-Yau manifold). The parameters controlling the internal geometry, i.e. the moduli, manifest as dynamical scalar fields in the $d$-dimensional EFT, yielding an action akin to Eq.~\eqref{sectEFT}. These fields are generically massless at tree level and thus pose phenomenological problems like mediating long-range fifth forces. Granting the moduli phenomenologically viable masses becomes an essential task for compactification model building and goes under the umbrella of moduli stabilisation mechanisms (see \cite{McAllister:2023vgy} and references therein).

The Swampland Distance Conjecture, in its most common form, can be presented as follows: 
\begin{conjecture}{Swampland Distance Conjecture \cite{Ooguri:2006in}}
    Consider the moduli space $\mathcal{M}$ of a gravitational EFT in $d\geq 4$ dimensions, parameterised by the vevs of massless scalar fields. Compared to the theory at some point $P\in \mathcal{M}$, the theory at another point $Q\in \mathcal{M}$ has an infinite tower of states, whose characteristic mass scales as
    \begin{equation}\label{eq:exponentialtower}
        m_\text{tower}(Q)\sim m_\text{tower}(P)e^{-\lambda_t\Delta_\phi}\,,
    \end{equation}
    where $\Delta_\phi$
    is the distance along a geodesic $\gamma$ in $\mathcal{M}$ connecting $P$ and $Q$, and $\lambda_t$ is an $\mathcal{O}(1)$ number measured in $d$-dimensional Planck units.
\end{conjecture}
Whilst infinite towers of states are far from common occurrence in the context of ordinary scalar-tensor theories, they appear quite naturally within string theory. One example is when integrating out compact extra dimensions in the presence of a bulk scalar field; each Fourier mode will itself look like a scalar field in lower dimensions, with mass being an integer multiple of the scale $m_\text{tower}$. These are commonly known as Kaluza-Klein (KK) towers. Another example is the excitations of the fundamental string itself, which also gives rise to an infinite tower of states. In fact, the Emergent String Conjecture \cite{Lee:2019wij}  advocates that infinite towers of states, when invoking the Swampland Distance Conjecture, will always fall into one of the two categories.

It is worth noting that the Swampland Distance Conjecture really contains two logically distinct pieces.
\begin{enumerate}
    \item (Existence) There exists at least one infinite tower of states.\footnote{From our previous paragraph, one may worry that requiring existence of infinite towers of states necessitates a strong devotion to string theory by default. This is not necessarily the case. It can be argued independently using tree-level graviton scattering amplitudes that UV-finiteness requires an infinite number of resonances \cite{Alonso:2019ptb, Huang:2022mdb}.}
    \item (Distance) The characteristic mass scale of at least one tower scales with the distance as stated above.
\end{enumerate}
Following the original proposal, significant effort was devoted to understanding the parameter $\lambda_t$ that governs the rate at which towers of states become light. Accumulating evidence and interplay with other proposals such as the Weak Gravity Conjecture suggested that $\lambda_t$ could not be any arbitrary order one parameter, but rather exhibited some universal features independent of the model in use. This led to the now well-established sharpened version of the Swampland Distance Conjecture:

\begin{conjecture}{Sharpened Distance Conjecture (SDC) \cite{Etheredge:2022opl}}
    In quantum gravity in $d\geq 4$ dimensions, any infinite distance limit in moduli space features at least one tower of states which satisfies the Swampland Distance Conjecture with
    \begin{equation}
        \lambda_t\geq \frac{1}{\sqrt{d-2}}\,.
    \end{equation}
\end{conjecture}
The formulation in Eq.~\eqref{eq:exponentialtower} is most reliable in asymptotic regions of moduli space --- namely, near infinite-distance limits where one has good analytic control of the theory. Away from these asymptotic regions, in the bulk interior of moduli space, explicit computation in string theory is often difficult or unavailable. One typically lacks an exact expression for $m_\text{tower}(\phi)$ or even the precise metric that enters the definition of $\Delta_\phi$. For this reason, more recent work tends to address the conjecture in more geometric terms. One introduces the so-called tower vector $\zeta$ (also sometimes called the scalar charge-to-mass vector), given by
\begin{align}\label{chargetomassvec}
    \zeta^i = - \mathsf{G}^{ij} \nabla_j \ln \left( \frac{m_\text{tower}}{\Mpld}\right)\,.
\end{align}
This vector captures the direction in moduli space along which the tower becomes light the fastest, and its norm encodes the rate at which the mass varies. The SDC thereby places a lower bound on its norm
\begin{align}
    \abs{\zeta} \geq \frac{1}{\sqrt{d-2}}\,.
\end{align} 
The main take-away of the SDC is that the rate of change is never `small': super-Planckian field excursions will always be met with an infinite tower of states becoming light, challenging the validity of the original EFT. The cutoff is perhaps the most direct indicator of this. In the presence of $\Nsp$ gravitationally coupled species, the strong coupling scale of gravity is no longer $\Mpld$, but lowered to (at most) the species scale \cite{Dvali:2007hz,Dvali:2007wp},
\begin{equation}
    \Lambda_\text{cutoff} \lesssim \Lsp = \frac{\Mpld}{\Nsp^\frac{1}{d-2}}\,.
\end{equation}
An \emph{increase} in the number of light gravitationally interacting degrees of freedom thereby \emph{decreases} the cutoff scale. This lends credence to the idea that there must be an analogue version of the SDC for the species scale $\Lambda_{\rm sp}$.
\begin{conjecture}{Species Scale Distance Conjecture (SSDC) \cite{vandeHeisteeg:2023ubh}}
    Consider a gravitational EFT in $d\geq 4$ dimensions whose UV cutoff is said to be bounded by the species scale ${\Lambda_\text{UV}\lesssim \Lsp}$. Compared to the theory at some $P\in\mathcal{M}$ in the moduli space, the species scale at $Q\in\mathcal{M}$ towards an infinite distance limit behaves as
    \begin{equation}
        \Lsp(Q)\sim \Lsp(P) e^{-\lambda_\text{sp} \Delta_\phi}\,,
    \end{equation}
    where $\lambda_\text{sp}$ is an $\mathcal{O}(1)$ number measured in $d$-dimensional Planck units, and obeys the bound
    \begin{equation}
        \lambda_\text{sp}\leq \frac{1}{\sqrt{d-2}}\,.
    \end{equation}
\end{conjecture}
Similar to the SDC, this conjecture is also phrased in terms of a vector $\mathcal{Z}$, called the species vector,
\begin{equation}
    \mathcal{Z}^i = -\mathsf{G}^{ij}\nabla_j \ln\left(\frac{\Lsp}{\Mpld}\right)\,.
\end{equation}
Its norm is subject to both an upper and lower bound, with the latter put forward soon after in \cite{Calderon-Infante:2023ler},
\begin{equation}\label{eq:speciesvectorconjecturebounds}
    \frac{1}{\sqrt{(d-1)(d-2)}} \leq \abs{\mathcal{Z}}  \leq \frac{1}{\sqrt{d-2}}\,.
\end{equation}
Again, the take-away is that the rate of change of $\Lsp$ is bounded, which has severe consequences for models relying on relatively large field excursions or energy scales comparable to $\Mpld$. Inflation (un)fortunately falls into both categories: demanding that the inflationary scale lies below $\Lsp$ at the end of inflation leads to a bound on the field excursion \cite{Scalisi:2018eaz}.
\begin{equation}
    H\lesssim \Lsp(\phi_\text{end}) \simeq \Lsp(\phi_\text{start})e^{-\lambda_\text{sp} \Delta_\phi} \lesssim \Mpld e^{-\lambda_\text{sp}\Delta_\phi}
\end{equation}
In $d = 4$ dimensions, this implies
\begin{equation}
\Delta_\phi \lesssim \sqrt{6} \ln \left( \frac{M_\text{Pl}}{H} \right)\,.
\end{equation}
Typical values for $H$ lie in the range $(10^{-5} - 10^{-3}) M_\text{Pl}$, which bounds $\Delta_\phi \lesssim 10$. Of course, even before one reaches this bound, the shifts in $\Lsp$ and $\Nsp$ can call into question the validity of inflationary models. An example is Starobinsky inflation \cite{Lust:2023zql}. There have also been attempts to bend the implied kinematics of these conjectures to the advantage of certain inflationary models \cite{Scalisi:2024jhq,Casas:2024jbw,Casas:2024oak,Cribiori:2025oek}. Without a doubt, the interplay between inflation and the Distance Conjectures is one of the most bright-lined links to study phenomenological consequences of the Swampland Programme; further inquiry remains relevant in anticipation of upcoming experiments which may test these claims.

Let us conclude this section by highlighting a relatively recent development that lies at the intersection of the SDC and SSDC, and has drawn considerable attention within the Swampland Programme. The convex hulls \cite{Etheredge:2022opl, Calderon-Infante:2020dhm} spanned by the tower- and species vectors of different towers of states have revealed a striking regularity: the inner products of these vectors equate to a universal constant in the vicinity of infinite-distance limits. This observation has motivated a so-called `universal pattern', conjectured to govern the behaviour of the lightest tower of states.

\begin{conjecture}{A ``Universal Pattern'' \cite{Castellano:2023stg, Castellano:2023jjt}}
    Asymptotically towards an infinite distance limit in the moduli space, 
    \begin{equation}
        \zeta \cdot \mathcal{Z}=\mathsf{G}^{ij}\frac{\nabla_i m_\text{tower}}{m_\text{tower}} \frac{\nabla_j \Lambda_\text{sp}}{\Lambda_\text{sp}}= \frac{1}{d-2}\,,
    \end{equation}
    where $m_\text{tower}$ is the the mass of the \emph{lightest} tower of states in this limit and $\nabla$ denotes the gradient in moduli space.
\end{conjecture}

\section{Frame Covariance and Field Space Geometry}
\label{sec:fieldspacegeometry}

In this section, we will now review the concept of field space geometry the way it has been studied in the context of scalar-tensor theories and frame covariance. One of the core ideas from this point of view is the intimate relation between Weyl transformations and unit transformations. We want to emphasise that the discussion surrounding frame covariance becomes particularly important for \emph{gravitational} EFTs, and we shall see that the conformal mode also plays a crucial part later in Section \ref{sec:augmentedfieldspace}. The spirit somewhat mimics that of the Swampland Programme: if one `turns off gravity', frame transformations become entirely trivial. In fact, let us ground our discussion by starting with an action very similar to Eq.~\eqref{sectEFT}. Consider the following generic $d$-dimensional scalar-tensor theory
\begin{equation}\label{eq:scalartensoraction}
    S = \int \dd^dx\ \sqrt{-g}\left[ \frac{f(\bphi)}{2}R[g] - \frac{1}{2}g^{\mu\nu}\mathsf{G}_{ij}(\bphi)\partial_\mu\phi^i\partial_\nu\phi^j - V(\bphi)\right]\,.
\end{equation}
Theories such as this appear in many contexts: Brans-Dicke theory \cite{PhysRev.124.925}, $f(R)$ gravity \cite{Sotiriou:2006hs}, (beyond) Horndeski theory \cite{Kobayashi:2019hrl}, quintessence \cite{Tsujikawa:2013fta}, and models with varying effective constants \cite{Barrow:2011kr}, among several others. It is common convention to say that the action in Eq.~\eqref{eq:scalartensoraction} is written in the \emph{Jordan} frame. This means that the scalars are non-minimally coupled to gravity, unlike Eq.~\eqref{sectEFT}, which is said to be in the \emph{Einstein} frame. For simplicity we will not be discussing the coupling to matter (although its presence can be incorporated in a frame-covariant formalism, see e.g. \cite{Jarv:2014hma,Kuusk:2015dda}). 

At first glance, one might propose that the theories captured by Eqs.~\eqref{sectEFT} and~\eqref{eq:scalartensoraction} are completely different in structure, even without the presence of a scalar potential; one contains non-minimally coupled scalars, the other does not. However, it has been long established that this is not the case. We see this when performing a Weyl transformation,\footnote{Somewhat confusingly, much of the cosmology literature refers to a local rescaling of the metric as a ``conformal transformation'' rather than a Weyl transformation. The two may look similar, but they are inherently different: conformal transformations are \emph{active coordinate transformations} while a Weyl transformation is a map directly on the metric itself. In this paper we will try to stay consistent with calling the rescaling of the metric a Weyl transformation.}
\begin{equation}\label{eq:Weyltransformation}
g_{\mu \nu}\longmapsto \hat g_{\mu\nu} =\Omega^2 g_{\mu\nu}\,.
\end{equation}
By a judicious choice of $\Omega$, one can cast the action in such a way that the scalars appear minimally coupled. In this sense, the Jordan and Einstein frames manifest as different ways of presenting the same underlying theory.

There has been a long-standing debate on whether the choice of frame has any physical bearing. This disagreement has been termed the \emph{frame problem}, and it has been an ongoing topic of discussion for many years \cite{Faraoni:1999hp,
    Capozziello:2006dj, 
    Catena:2006bd, 
    Faraoni:2006fx, 
    Capozziello:2010sc, 
    Weenink:2010rr,
    Gong:2011qe, 
    Quiros:2011iv, 
    Kubota:2011re, 
    White:2012ya, 
    Quiros:2012rnn, 
    Calmet:2012eq, 
    Steinwachs:2013tr, 
    Prokopec:2013zya, 
    White:2013ufa, 
    Postma:2014vaa, 
    Karamitsos:2017elm, 
    Ohta:2017trn, 
    Ruf:2017xon,  
Falls:2018olk}. There is broad consensus that the frame problem is resolved at the classical level, i.e. it has been demonstrated that the choice of frame does not affect the underlying physics of a theory, even if subtleties arise when matter couplings or radiative corrections are involved. Still, the physical meaning of a Weyl transformation has been much discussed in the literature in the context of the frame problem. Allow us therefore to digress slightly to unpack its intimate relationship with units and unit transformations.

\subsection{Weyl Transformations and Unit Transformations} \label{sec:unittransformations}
The choice of units presents a juxtaposition of both importance and unimportance. The importance of selecting a system of units (one for length, mass, time, etc.) lies in its necessity in communicating the magnitude of dimensionful quantities, which are number-unit pairs, e.g. $m_e=\SI{9.11e-31}{\kilogram}$. However, at the same time, the unimportance of units is evident when we consider the freedom to choose any system of units we want. Indeed, the two statements $m_e=\SI{9.11e-31}{\kilogram}$ and $m_e=\SI{0.51}{\mega\electronvolt\slash c^2}$ convey the exact same message as long as we know how to convert between $\SI{}{\kilogram}$ and $\SI{}{\mega\electronvolt\slash c^2}$. It should thus be uncontroversial that the numerand of a dimensionful quantity depends entirely on the human construct that is our choice of units. Dimension\emph{less} quantities, on the other hand, possess a much stronger gravitas. Take the fine structure constant or the ratio between the mass of a proton and an electron. Two physicists working in different unit systems (e.g. SI and Imperial) would arrive at the exact same answer without needing translation between conventions.

When we turn on gravity, a new subtlety arises because the line element $\dd s^2$ itself acts as a natural spacetime ruler. In special relativity, we know that observers linked by Lorentz transformations will agree on spacetime intervals and hence they may consistently share the same spacetime ruler. However, under a Weyl transformation $\dd s^2 \mapsto \dd\tilde{s}^2 = \Omega^2 \dd s^2$, the line element is no longer invariant, and we find ourselves invited to use an exotic system of \emph{spacetime-dependent units}. 

The consequence becomes sharp once we consider the effect on the measurement of some dimensionful quantity $Q$. Operationally, we compare $Q$ to some choice of reference scale $\mathrm{M}$ (our ruler) by constructing the dimensionless 
\begin{equation}
    \Pi = \frac{Q}{\mathrm{M}^{[Q]}}\,,
\end{equation}
where $[Q]$ is the ``mass'' or ``engineering'' dimension of $Q$, $[\mathrm{M}]=1$, and we work in some natural set of units so that mass, inverse length, inverse time, and so on share the same dimensionality.

Now suppose the metric undergoes a Weyl transformation like Eq.~\eqref{eq:Weyltransformation}. One can take two equivalent, yet subtly different, viewpoints on how this affects dimensionful quantities:
\begin{enumerate}[label=\roman*)]
    \item \textbf{Unit-first point of view (frame covariance):} regard Eq.~\eqref{eq:Weyltransformation} as redefining the spacetime metric $g_{\mu\nu}\mapsto \hat{g}_{\mu\nu}$ and thus its associated ruler $\mathrm{M}\mapsto \hat{\mathrm{M}}=\Omega^{-1}\mathrm{M}$. Rescaling the ruler means that quantities $Q$ transform as
    \begin{equation}
        Q\longmapsto \hat{Q} = \Omega^{-[Q]}Q\,,
    \end{equation}
    leaving the dimensionless ratio $\Pi$ invariant
    \begin{equation}
        \Pi = \frac{Q}{\mathrm{M}^{[Q]}} \longmapsto \frac{\hat{Q}}{\hat{\mathrm{M}}^{[Q]}}=\frac{\Omega^{-[Q]}Q}{(\Omega^{-1}\mathrm{M}
        )^{[Q]}} =\Pi\,.
    \end{equation}
    From this viewpoint it is clear that a Weyl transformation is simply isomorphic to a local change of units. This is the picture adopted in most literature on frame covariance.
    \item \textbf{Metric-first point of view (field excursions):} a complementary description is to perform the Weyl transformation but insist on working with the original metric $g_{\mu\nu}$ and fixed ruler $\mathrm{M}$ (for example $d$-dimensional Planck units). This picture is more commonly adopted in a Swampland context, where field excursions effectively induce Weyl transformations via Weyl rescaling factors, but one has in mind to always cast the action in terms of the same background metric $g_{\mu\nu}$ and unit $\Mpld$ (more on this in Section~\ref{sec:distanceconjecture}).  The result is that quantities transform as $Q\mapsto \Omega^{[Q]}Q$ and the observable $\Pi$ thus also appears to scale by
    \begin{equation}\label{physicalchange}
        \Pi = \frac{Q}{\mathrm{M}^{[Q]}}\longmapsto \frac{\Omega^{[Q]}Q}{\mathrm{M}^{[Q]}} = \Omega^{[Q]}\Pi\,.
    \end{equation}
    However, this ``physical effect'' is purely a remnant of the fact that we have insisted on using the same ruler in-between two different conformal frames. By performing the appropriate unit transformation $\mathrm{M}\mapsto \Omega\mathrm{M}$ in combination with the Weyl transformation
    \begin{equation}\label{unitchange}
        \Pi = \frac{Q}{\mathrm{M}^{[Q]}} \longmapsto \frac{\Omega^{[Q]}Q}{(\Omega\mathrm{M})^{[Q]}} = \Pi
    \end{equation}
    dimensionless ratios are left invariant. The main point is that, in this picture, unit transformations may always \emph{undo} the effect of Weyl transformations.
\end{enumerate}

The two views presented above are most of all a matter of taste and at the end of the day isomorphic to one another. Whether we view a Weyl transformation as a unit transformation or a rescaling of every dimensionful quantity in the universe, the result is the same: the transformation is not observable. This is fundamentally because it is impossible to measure dimensionful quantities. Any measurement of such a quantity necessarily involves comparing it against another commensurable fixed physical scale, e.g. the length of the King's foot. We can therefore only really measure dimension\emph{less} quantities. In this sense, our  `choice of units' is really not much different from a `choice of gauge'.

The idea that physics is fundamentally dimensionless has a long history and is explicitly formalised through the Buckingham-$\pi$ theorem in dimensional analysis. It states that any physically meaningful statement (i.e.~independent of human convention) of the form $f(Q_1,\dots,Q_N) = 0$, where $Q_1,\dots,Q_N$ are quantities collectively involving $n$ base dimensions,  can always be recast as some equivalent statement $F(\pi_1,\dots,\pi_{N-n})=0$ in terms of $N-n$ dimensionless quantities $\pi_1,\dots,\pi_{N-n}$.\footnote{By base dimensions we mean a choice of distinct dimensionalities, such as the typical engineering dimensions $\{\mathrm{M}, \mathrm{L}, \mathrm{T}\}$, and so on. However, we can be more general in this notion, see \cite{Jacobs:2024} for a more careful discussion.} In other words, the Buckingham-$\pi$ theorem tells us that physics is fundamentally dimensionless. This fact is perhaps not widely acknowledged or appreciated within theoretical physics today: indeed, most textbooks and literature will adopt natural units with everything set to $1$ to circumvent the hassle of keeping track of dimensionalities. 

It is not difficult to see that a global change of units, with a constant conversion factor, does not affect physics. After all, even when considering dimensionful statements like $f(Q_1,\dots,Q_N)=0$, the conversion factors drop out as long as said relations are dimensionally homogeneous and consistent. In physics, promoting global symmetries to local ones has been a catalyst for many important insights, and unit transformations are no exception. When we promote the change of units to be local, derivatives of dimensionful quantities become sensitive to the choice of units, since the ordinary derivative no longer commutes with $\Omega$. This motivates the definition of a unit-covariant derivative,
\begin{equation}\label{unitcovder}
\mathrm{D}_\mu Q \equiv \partial_\mu Q - [Q] \frac{\partial_\mu\mathrm{M}}{\mathrm{M}} Q,
\end{equation}
where $\mathrm{M}$ is an (arbitrary) unit of $[\mathrm{M}]=1$.\footnote{If we were not working in natural units, such a unit-covariant derivative would require additional terms, one for each base unit.} Put simply, it is just the gauge-covariant derivative associated with local unit transformations. Of course, relations written purely in terms of dimensionless quantities and their derivatives are sufficient to express physics in a unit-invariant way. But when dealing with dimensionful quantities, promoting $\partial \to \mathrm{D}$ becomes necessary when comparing results before and after a local Weyl or unit transformation.

Let us close on a somewhat light-hearted analogy that illustrates our discussion of Weyl transformations. Imagine we are visiting a funhouse featuring multiple distorted mirrors, bringing along of course our trusty ruler. Looking to one of the mirrors we might see ourselves stretched to the height of a basketball player. In another, we have shrunk to half our size and are ready to venture across the Shire. With ruler in hand we can confirm this apparent rescaling: keeping it next to the mirror and using it as a unit, we really are taller or shorter in the mirror image. However, this is nothing but a consequence of us insisting on using the same reference scale --- fundamentally, our heights have not actually changed at all. The way to reconcile this fact with observations is that, when holding the ruler opposite the mirror, the ruler \emph{itself} is rescaled in the mirror image. Separate ratios of lengths inside and outside of the mirror do not give us any physical information on its own. More specifically, if we were to stack rulers from head to toe, even if the curvature of the mirror is such that the ruler changes length as it travels up and down (corresponding to a local change of units), both ourselves and our mirror counterparts will agree on the dimensionless quantity that is our height to ruler ratio. 
 
\subsection{Frame Covariance and the Field Space Metric}

If Weyl transformations are as physically meaningless as the choice of units (inasmuch a careful selection of units can undo the effects of a Weyl transformation), it should follow that theories may be cast in a way that is agnostic to the frame that they are expressed in --- analogous to dimensionless quantities which are agnostic to the system of units one may fancy. It turns out that this structure is deeply geometric. 

Given an EFT like Eq.~\eqref{eq:scalartensoraction}, it is very common in the literature to read off from the action the kinetic function $\mathsf{G}_{ij}$ and dub it the field space metric. There is merit to this convention (and it is indeed a convention), since many established results follow straightforwardly from such a choice. Originally, the geometric approach to the field space was motivated by searching for an ultralocal connection to the configuration space in order to define the unique effective action \cite{Vilkovisky:1984st}. But, since the resulting connection is Levi-Civita, formulating the field space by imposing the form of the metric is arguably more fundamental. In more complicated theories where the metric is not immediately obvious to identify, one requires of it three properties: (i) it is a symmetric rank-2 tensor, (ii) Euclidean for a canonically normalized theory, and (iii) it can be fully determined from the classical action \cite{GrootNibbelink:2000vx,GrootNibbelink:2001qt, vanTent:2003mn}. For \emph{minimally} coupled scalar-tensor theories, this is an easy task. The metric can be immediately read off and identified with the kinetic function~$\mathsf{G}_{ij}$.\footnote{For theories such  as ones involving higher-order derivatives, the metric cannot be straightforwardly read off the action, but it is possible to extend the definition of field-space metric such that the above requirements are still met~\cite{Finn:2019aip}.}
 
In the presence of a \emph{non-minimal} coupling to gravity, we might expect that the same holds true. However, as we have argued above, we seek and expect there to be a notion of field space metric that is insensitive to Weyl transformations. This makes the kinetic term inappropriate, because it does not transform covariantly under a frame transformation. Indeed we find that (in the unit-first point of view) the action becomes\footnote{Here the conformal factor depends on spacetime only through the fields $\Omega(x) \equiv \Omega(\bphi(x))$.} 
\begin{equation}\label{actiongen}
    S = \int \dd^dx\ \sqrt{-\hat{g}}\left[ \frac{\hat{f}(\bphi)}{2}\hat{R}[\hat{g}] - \frac{1}{2}\hat{g}^{\mu\nu}\hat{\mathsf{G}}_{ij}(\bphi)\partial_\mu\phi^i\partial_\nu\phi^j - \hat{V}(\bphi)\right]\,,
\end{equation}
where $\hat{R}[\hat{g}]$ is the Ricci scalar defined in terms of the transformed metric and the transformed functions are given by
\begin{equation}\label{transrules}
\begin{aligned}
\hat{f} &= \Omega^{-(d-2)}f\,,\\ 
{\mathsf{\hat{G}}}_{ij} &=   \Omega^{-(d-2)}\Big\{ \mathsf{G}_{ij} + (d-1)\Big(   f_{,i} (\ln \Omega)_{,j} 
  + f_{,j}(\ln \Omega)_{,i}\Big)  - (d-1)(d-2)f(\ln \Omega)_{,i}  (\ln \Omega)_{,j} 
   \Big\}\,,
\\
 \hat{V} &= \Omega^{-d}V\,.
\end{aligned}
\end{equation}
The transformation makes it clear that $\mathsf{G}_{ij}$ should \emph{not} be taken as the field space metric in a frame-covariant sense; it does not transform as a simple tensor under Weyl transformations. Nonetheless, using the transformation rules, one can construct a combination that \emph{is} Weyl invariant, 
\begin{align}\label{fieldspacemetriceinstein}
\mathcal{G}_{ij} = M_\mathsf{E}^{d-2}\left(\frac{\mathsf{G}_{ij}}{f} + \frac{d-1}{d-2} \frac{f_{,i} f_{,j}} {f^2}\right).
\end{align}
The prefactor of $M_\mathsf{E}^{d-2}$ is required for dimensional consistency. In Section \ref{sec:augmentedfieldspace} we shall demonstrate that it coincides with the induced metric on Einstein frame:\footnote{For single-field models, the term `Einstein frame' is often reserved for the form of the action with both minimal coupling and canonical kinetic term. In multi-field theories the kinetic term can only be made strictly canonical if the field space is flat. So with that in mind we use Einstein frame to just mean the frame in which the scalars are minimally coupled to gravity.} choosing ${\Omega^{d-2} = f/M_\mathsf{E}^{d-2}}$ so that the transformed action is minimally coupled with $M_\mathsf{E}$ the Einstein frame (reduced) Planck mass, the kinetic term of scalars is precisely $\mathcal{G}_{ij}\partial\phi^i\partial\phi^j$. Therefore, $\mathcal{G}_{ij}$ is a natural candidate for the field space metric associated with the non-minimal action \eqref{eq:scalartensoraction}.

By identifying the correct metric, the field space can be properly formalised as a Riemannian manifold with all the usual accoutrements of differential geometry: indices are raised and lowered using $\mathcal{G}_{ij}$ and its inverse, the line element is $\mathcal{G}_{ij} \dd \phi^i \dd\phi^j$, curvature tensors can be defined in the usual way, and parallel transport is encoded in the field space covariant derivative
\begin{equation}\label{fieldspacecovder}
D_i X^a \equiv X^a_{,j} + \Gamma^a_{ij} X^j,
\end{equation}
with the usual Levi--Civita connection
\begin{equation}
\Gamma^k_{ij} = \frac{1}{2} \mathcal{G}^{kl}\left( \mathcal{G}_{lj,i}  +  \mathcal{G}_{il,j} -\mathcal{G}_{ij,l} \right)\,.
\end{equation}
Under a field reparameterisation $\phi^i\to \hat{\phi}^{\hat{i}}(\bphi)$ with Jacobian
\begin{equation}
\tensor{J}{^{\hat{\imath}}_i} \equiv \frac{\dd \hat{\phi}^{\hat{\imath}}}{\dd \phi^i}
\end{equation}
the metric transforms as a standard $(0,2)$-tensor
\begin{equation}
\mathcal{G}_{ij}\longmapsto \hat{\mathcal{G}}_{\hat{\imath}\hat{\jmath}} =  \mathcal{G}_{ij} \tensor{J}{^i_{\hat{\imath}}} \tensor{J}{^j_{\hat{\jmath}}} \,.
\end{equation}
In the frame covariance literature, a \emph{frame transformation} refers to a combined Weyl transformation followed by a field reparameterisation. It is useful to identify \emph{frame-covariant} quantities as ones that transform as tensor densities under frame transformations
\begin{equation}
\tensor{X}{^{i_1i_2\cdots}_{j_1j_2\cdots}} \longmapsto \tensor{\hat{X}}{^{\hat{\imath}_1\hat{\imath}_2\cdots}_{\hat{\jmath}_1\hat{\jmath}_2\cdots}} = \Omega^{-c_X} (\tensor{J}{^{\hat{\imath}_1}_{i_1}} \tensor{J}{^{\hat{\imath}_2}_{i_2}} \cdots) 
\tensor{X}{^{i_1i_2\cdots}_{j_1j_2\cdots}} (\tensor{J}{^{j_1}_{\hat{\jmath}_1}} \tensor{J}{^{j_2}_{\hat{\jmath}_2}} \cdots)
\end{equation}
where $c_X$ is the conformal weight of the quantity. Because of the explicit $\Omega$-factor, the field space covariant derivative defined in Eq.~\eqref{fieldspacecovder} acting on $X$ does not transform covariantly. Using the fact that $f(\bphi)$ has conformal weight $d-2$, one can define a fully frame-covariant derivative
\begin{equation}\label{framecovder}
    \mathcal{D}_i X = D_i X - \frac{c_X}{d-2}\frac{f_{,i}}{f} X
\end{equation}
whose index structure is determined by that of $X$. A more special case of this derivative appears in \cite{Postma:2014vaa}. This derivative is the natural analogue --- in field space --- of the unit-covariant derivative in Eq.~\eqref{unitcovder}. Both encode how quantities with non-trivial weights change under local rescalings.

Finally, we briefly remark that these ideas extend to configuration space, where the fields $\phi^i(x)$ themselves serve as coordinates. The configuration space is an infinite-dimensional Riemannian manifold with an ultralocal metric where ordinary derivatives are promoted to functional derivatives and the frame-covariant approach yields a unique reparameterisation invariant effective action. For instance, the usual one-loop effective action (which is not reparameterisation invariant \cite{vilkovisky1984gospel,Finn:2019aip}) can be extended to give a fully covariant expression \cite{Vilkovisky:1984st,Rebhan:1986wp,Burgess:1987zi,Ellicott:1987ir}
\begin{equation}
    \Gamma_1[ \phi ] = \frac{i}{2}  \text{tr} \ln  \Big( \nabla^i \nabla_i S[ \phi] \Big)\,,
\end{equation}
where $\nabla_i$ denotes the covariant functional derivative compatible with the configuration space metric $\mathcal{G}_{ij} \delta(\phi^i - \phi^j)$. This approach can also be applied to covariantly describe the superspace manifold in quantum gravity. 

\section{Field Space Geometry of Gravitational EFTs
}\label{sec:augmentedfieldspace}

After providing background for the Swampland Programme, some of its Distance Conjectures, as well as frame invariance and covariance, we are now ready to lay the groundwork that will allow us to draw connections between these seemingly disparate topics. The key task will be to formulate a frame-covariant understanding of the field space geometry describing gravitationally interacting scalars from a metric-first point of view (cf. Section~\ref{sec:unittransformations}). It is fair to say that such a task has been somewhat overlooked from the perspective of the Swampland Programme. The application is however immediate: if Einstein, Jordan, and any other conformal frame ought to be physically selfsame, then we should expect there to be a sense in which the field space geometry is `generally covariant under frame transformations'. Moreover, we discussed and demonstrated in the previous section that one may be too hasty in identifying the function $\mathsf{G}_{ij}(\bphi)$ appearing in the scalar kinetic term as the metric on field space. Eq. \eqref{transrules} shows that $\mathsf{G}_{ij}(\bphi)$ transforms non-trivially (i.e. not just up to a conformal factor) under Weyl transformations. Notions which involve a choice of metric, such as the distance in field space, thus give off the impression that they explicitly depend on a choice of frame, going completely against the fact that physics should be frame-independent. 

A second point of contention concerns the reliance on units in the standard formulations of the Distance Conjectures. Both the SDC and SSDC (and, in fact, essentially all Distance Conjectures) are phrased in `$d$-dimensional Planck units'. This is conceptually puzzling. After all, physical statements, as we discussed in Section \ref{sec:fieldspacegeometry}, are expected to be unit-independent. If Swampland Conjectures in particular are meant to capture deep constraints concerning UV completion in quantum gravity, why do they depend on a choice of measuring stick? If one redefines, for instance, the species vector $\mathcal{Z}$ in units of the species scale itself, the vector is trivial by construction, and the SSDC disappears. At the very least, this suggests that some essential physical context is missing, or there is superfluity obscuring the true content of these statements. It becomes even more curious that $d$-dimensional Planck units turn out to play an important role. Jacobs argues in \cite{Jacobs:2025kdw} that the physical significance of the Planck scale for quantum gravity is largely guided by questionable heuristics. Even within the Swampland Programme itself we do not typically regard $\Mpld$ as an obtainable energy scale --- the species scale $\Lsp$ is generically \emph{below} $\Mpld$ and thus, physically speaking, much more relevant. Working in units of an effectively insignificant scale then appears, at best, uneconomical.

Taken together, these considerations motivate a genuinely frame- and unit-covariant formulation of field space geometry as a central prerequisite for phrasing, applying, and scrutinising the Distance Conjectures in an unambiguous way.

\subsection{The Frame-Augmented Field Space}
The transformed kinetic function in Eq. \eqref{transrules} is derived under the assumption that the conformal factor $\Omega$ is a function of the fields $\phi^i$. For example, by choosing $\Omega^{d-2} \propto f(\bphi)^{-1}$, which corresponds to going to Einstein frame, one obtains Eq. \eqref{fieldspacemetriceinstein}. A choice of frame is thereby seeded by a choice of function $\Omega(\bphi)$, and the set of all such consistent functions gives rise to the full assortment of conformal frames. In considering all frames \emph{simultaneously}, one therefore has to take a step back and \emph{not} (yet) specify the form of $\Omega(\bphi)$. It is indeed possible to extract the conformal factor $\Omega$ from the spacetime metric and treat it on an equal footing with the $N$ scalar fields $\phi^i$. The resulting action now admits an $(N+1)$-dimensional field space, parameterised by the original fields $\phi^i$ alongside $\Omega$. This means that imposing a particular conformal frame by setting $\Omega = \Omega(\bphi)$ carries a geometric interpretation: it specifies a hypersurface in this augmented space! More precisely, this $(N+1)$-dimensional field space contains all possible conformal frames (and therefore all possible systems of units, even spacetime-dependent ones) as hypersurfaces of codimension one. For that reason, we find it fitting to dub this higher-dimensional space the \emph{frame-augmented field space}, or \emph{augmented field space} for brevity. This approach of treating the conformal factor as a field in its own right \cite{Bamber:2022eoy} has received some attention in the context of scattering amplitudes \cite{He:2023fko}, but not in the context of the Swampland Programme.

Consider again the general non-minimally coupled multi-scalar gravitational EFT in Eq.~\eqref{eq:scalartensoraction} and perform a Weyl transformation $g_{\mu\nu}\mapsto \Omega^2 g_{\mu\nu}$. Thanks to how the Ricci scalar transforms, the action acquires ``kinetic terms'' involving $\Omega$ (this is exactly why the transformation rule for the field space metric in Eq. \eqref{transrules} is not simply an overall factor). Adopting the metric-first point of view for Weyl transformations, one readily finds
\begin{equation}
	S = \int \dd^dx\sqrt{-g}\left[\frac{\Omega^{d-2}f(\bphi)}{2}R[g] - \frac{1}{2}\widehat{\mathsf{G}}_{IJ}(\bPhi) g^{\mu\nu} \partial_\mu\Phi^I \partial_\nu\Phi^J - \Omega^d V(\bphi)\right]\,.
\end{equation}
Here we have defined $\Phi^I=(\omega,\phi^i)$ and $\omega = \ln\Omega$ as the coordinates spanning the augmented field space, which we shall denote as $\mathcal{M}^\omega$. Using $\omega$ and not $\Omega$ as a coordinate turns out to be a convenient choice, but the string theorist may intuitively think of it as the $d$-dimensional dilaton --- more on this in Section \ref{sec:distanceconjecture}. We can read off the associated metric on $\mathcal{M}^\omega$ from the generalised kinetic term,
\begin{equation}\label{eq:augmentedmetric}
	\widehat{\mathsf{G}}_{IJ}(\bPhi) = \Omega^{d-2}\begin{pmatrix}
		-(d-1)(d-2)f(\bphi) & -(d-1)f_{,j}(\bphi)\\
		-(d-1)f_{,i}(\bphi) & \mathsf{G}_{ij}(\bphi)
	\end{pmatrix}\,,
\end{equation}
where $\mathsf{G}_{ij}(\bphi)$ is the original kinetic prefactor in Eq. \eqref{eq:scalartensoraction}.

Allow us to briefly digress in order to comment on conventions. So far, we have chosen to keep the field space metric dimensionful, with $[\widehat{\mathsf{G}}_{IJ}]=d-2$, and the fields $\Phi^I$ dimensionless. This is perhaps contrary to the more common convention found in the literature, where the metric is dimensionless and fields are dimensionful, i.e. $[\mathsf{G}_{IJ}]=0$ and $\widehat{\Phi}^I \equiv \mathrm{M}^\frac{d-2}{2}\Phi^I$, with $[\widehat{\Phi}^I] = \frac{d-2}{2}$ and $[\mathrm{M}]=1$ some unit. The two conventions are indeed equivalent once we promote all standard derivatives to their unit-covariant counterparts, $\partial \mapsto \DD$, since
\begin{equation}
    \mathsf{G}_{IJ} \DD_\mu \widehat{\Phi}^I \DD^\mu \widehat{\Phi}^J = \mathrm{M}^{d-2}\mathsf{G}_{IJ} \partial_\mu \Phi^I\partial^\mu \Phi^J = \widehat{\mathsf{G}}_{IJ} \partial_\mu \Phi^I \partial^\mu \Phi^J\,.
\end{equation}
In most of the literature surrounding the Swampland Programme, scalar fields are treated as dimensionful, meaning that the geometry on the associated field space is governed by the \emph{dimensionless} metric. For that reason, for the remainder of this paper, we will mostly work in terms of
\begin{equation}\label{eq:augmentedmetricdimless}
    \mathsf{G}_{IJ}(\bPhi) \equiv \frac{1}{\mathrm{M}(\bPhi)^{d-2}}\widehat{\mathsf{G}}_{IJ}(\bPhi) 
\end{equation}
where $\mathrm{M}$ is a choice of unit $[\mathrm{M}]=1$ that can be any function of $\Phi^I$. We shall see later how different choices of units impact questions relying on the geometry. 

Coming back on track, let us detail some features of the augmented field space metric. First of all, it is manifestly covariant under Weyl transformations, since it transforms as $\mathsf{G}_{IJ}\mapsto \Omega^{d-2}\mathsf{G}_{IJ}$. This is unlike $\mathsf{G}_{ij}(\bphi)$, which, as we observed already, picks up additional terms. By isolating the conformal factor, we have in a sense already performed a Weyl transformation, justifying why $\mathsf{G}_{IJ}$ will transform by a simple rescaling. In the same way that the scalar potential $V(\bphi)$ is a representative for the potential $\Omega^dV(\bphi)$ in all related conformal frames, the augmented field space metric $\mathsf{G}_{IJ}$ is sufficient in order to specify the non-minimal and kinetic couplings in \emph{any} frame. In addition, $\mathsf{G}_{IJ}$ is by construction covariant under unit transformations, with $\mathrm{M}\mapsto \Omega\mathrm{M}$ inducing $\mathsf{G}_{IJ}\mapsto \Omega^{-(d-2)}\mathsf{G}_{IJ}$. In the same manner as Eq. \eqref{unitchange}, a simultaneous Weyl- and unit transformation leaves the metric invariant, as to be expected from a dimensionless ratio of dimensionful quantities.

Another important property is that $\mathsf{G}_{IJ}$ corresponds to a \emph{Lorentzian} metric on $\mathcal{M}^\omega$ (this is provided we make some reasonable assumptions, such as $f(\bphi)>0$ and that $\mathsf{G}_{ij}$ has a positive signature). In particular, the conformal factor $\omega$ defines the ``timelike'' direction in the augmented field space. This is not an unknown fact: Euclidean quantum gravity famously suffers from the so-called `conformal factor problem', where the Euclidean path integral becomes unbounded due to the conformal mode having the ``wrong sign'' in its kinetic term \cite{Gibbons:1978ac}. Still, $\omega$ is a fundamentally unphysical degree of freedom as its presence can be undone by an appropriate unit transformation.

Finally, given that $\mathsf{G}_{IJ}$ is the appropriate metric on $\mathcal{M}^\omega$, conformal frames receive a natural geometric interpretation as codimension one (spacelike) submanifolds $\mathsf{\Sigma}\subset \mathcal{M}^\omega$, specified by the implicit equation $\Omega = \Omega(\bphi)$. The induced metric on this submanifold is
\begin{equation}
	\mathsf{G}_{ij}^\mathsf{\Sigma}(\bphi) = \left(\frac{\Omega(\bphi)}{\mathrm{M}(\bphi)}\right)^{d-2} \left\{\mathsf{G}_{ij}(\bphi)-(d-1)\frac{f_{,i}\Omega_{,j} + f_{,j}\Omega_{,i}}{\Omega(\bphi)} - (d-1)(d-2)f(\bphi) \frac{\Omega_{,i}\Omega_{,j}}{\Omega^2(\bphi)}\right\}\,,
\end{equation}
which, up to conventions, precisely matches the transformation rule in Eq.~\eqref{transrules} for the original kinetic function $\mathsf{G}_{ij}(\bphi)$. In other words, the field space metric for a particular conformal frame is just the \emph{induced metric} on the associated hypersurface $\mathsf{\Sigma}$ embedded in augmented field space.

The above motivates the definition of the Einstein frame (denoted by $\mathsf{\Sigma}=\mathsf{E}$) as the hypersurface where the scalars become minimally coupled to gravity, imposed by the implicit function 
\begin{equation}\label{eq:Einsteinframecondition}
	\mathsf{E}\colon\qquad  M_\mathsf{E}^{d-2} = \Omega^{d-2}f(\bphi)\,.
\end{equation}
Here $M_\mathsf{E}$ is any field-independent quantity with mass dimension $[M_\mathsf{E}]=1$. Therefore, the  induced field-space metric on Einstein frame reads
\begin{equation}\label{eq:dimfulaugmetric}
	\mathsf{G}_{ij}^\mathsf{E}(\bphi) = \left(\frac{M_\mathsf{E}}{\mathrm{M}}\right)^{d-2}\left(\frac{\mathsf{G}_{ij}(\bphi)}{f} + \frac{d-1}{d-2}\frac{f_{,i}f_{,j}}{f^2}\right)\,,
\end{equation}
which again is just Eq.~\eqref{fieldspacemetriceinstein} save for an overall constant that serves to make the induced field-space metric dimensionless. One can usually think of the constant $M_\mathsf{E}$ as the effective Planck mass or $d$-dimensional Planck mass.\footnote{The identification $M_\mathsf{E}\leftrightarrow \Mpld$ is quite intuitive in this case, however the story is much more subtle in the context of dimensional reduction, such as compactifications. We discuss in more detail the difference between $M_{\mathsf{E}}$ and $\Mpld$ in Section \ref{sec:distanceconjecture}.}

\begin{figure}[ht!]
	\centering
	\includegraphics[scale=0.6]{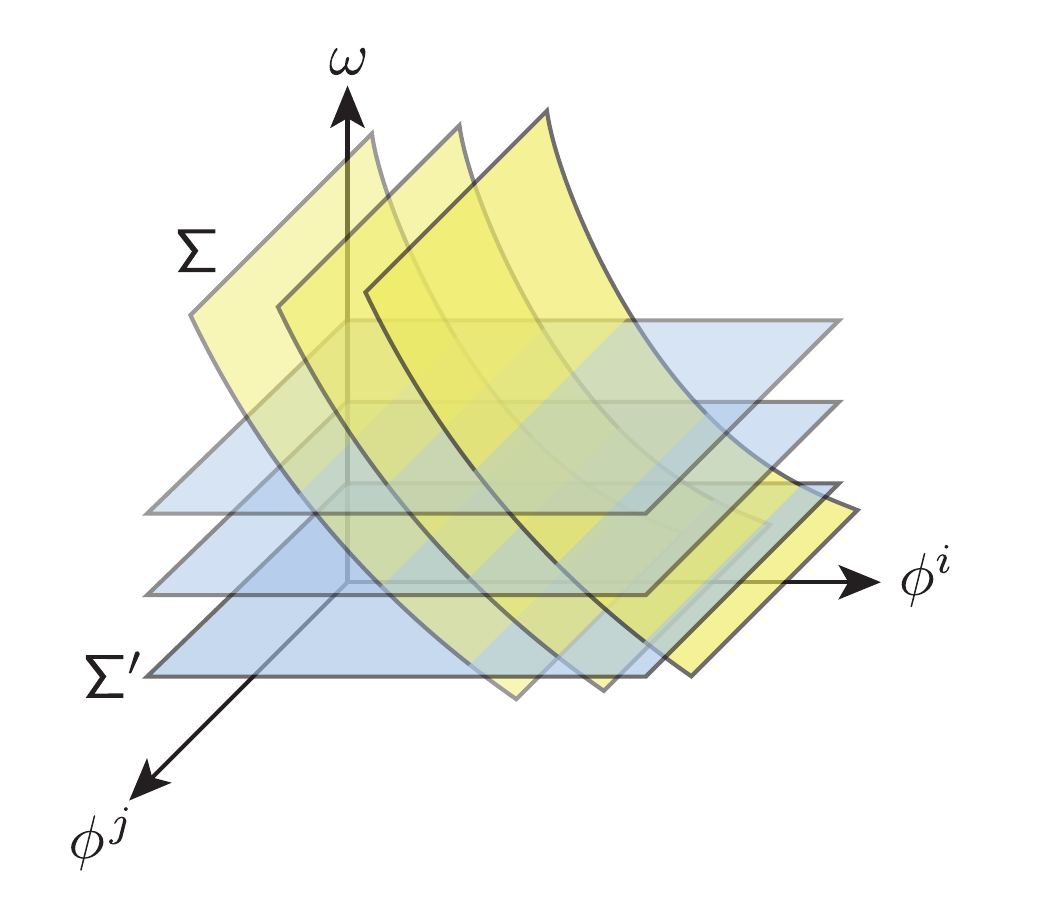}
	\caption{Illustration of the augmented field space $\mathcal{M}^\omega$ and different conformal frames $\mathsf{\Sigma},\mathsf{\Sigma}'$ as codimension one hypersurfaces. Every hypersurface gives rise to a foliation of frames related by constant Weyl transformations. A frame transformation $\mathcal{F}$ corresponds to a map sending leaves to leaves between different foliations.}
	\label{fig:augmentedfieldspace}
\end{figure}

In the literature it is incredibly common to encounter phrases referring to ``\emph{the} Einstein frame'', somehow implying that Einstein frame is uniquely defined. This, however, can be somewhat misleading. In fact, the definition of Einstein frame in Eq.~\eqref{eq:Einsteinframecondition} suggests that there exists --- not just one --- but a \emph{continuum} of Einstein frames, corresponding to different values of $M_\mathsf{E}$. This detail is rarely emphasised. From the perspective of the augmented field space, it is clear that this continuum of Einstein frames forms a one-parameter family of non-intersecting hypersurfaces $\{\mathsf{E}(M_\mathsf{E})\}$ that together span $\mathcal{M}^\omega$. In other words, the continuum of Einstein frames is nothing but a \emph{foliation} of $\mathcal{M}^\omega$. This is illustrated in Figure \ref{fig:augmentedfieldspace}.

Now, despite the above remark about the continuum of frames, we should still acknowledge that simply ignoring the fact has never gotten anyone into trouble: for any practical purposes, it is always possible to equate $M_\mathsf{E}$ with an arbitrary constant of interest or convenience (often the Planck mass). Moreover, different conformal frames are physically equivalent reparameterisations of the same physics, so the distinction between different Einstein frames therefore seems entirely superfluous. While all this is correct, we would like to underscore that the \emph{underlying reason} one can ignore the continuum of frames is not necessarily a given fact. Rather, it is deeply rooted in the discussion of unit transformations in Section~\ref{sec:unittransformations}.  

Recall that our choice of frame (i.e. a hypersurface $\mathsf{\Sigma}\subset \mathcal{M}^\omega$) and choice of unit (i.e. some dimensionful scale $\mathrm{M}$) are \emph{a priori} entirely independent of one another. A popular combination is to work ``in Einstein frame'' and ``in Planck units'', which is to say that we have made a simultaneous choice of hypersurface $\mathsf{E}(M_\mathsf{E})$ as well as a base unit $\mathrm{M}$ such that $M_\mathsf{E}/\mathrm{M}=1$. Of course, such a combination is not unique: one could have chosen any \emph{other} Einstein frame $\mathsf{E}(M_\mathsf{E}')$ with $M_\mathsf{E}'=\lambda M_\mathsf{E}$ and $\lambda >0$. These two choices can be related by a constant Weyl transformation. Notwithstanding, the two frames return the exact same (dimensionless) observables
\emph{provided} we work in units of $\mathrm{M}'=\lambda\mathrm{M}$, which is nothing but a constant unit transformation. 

Geometrically, constant Weyl transformations map leaves to leaves within the same foliation of $\mathcal{M}^\omega$, inducing a constant rescaling of every dimensionful quantity. This is in turn undone by a corresponding unit transformation, cf.~Eqs.~\eqref{physicalchange} and \eqref{unitchange}, and is the fundamental reason that we can neglect the continuum of Einstein frames (or any other frame): unit transformations can undo the effect of Weyl transformations on dimensionless observables. It suffices to consider just one leaf in any foliation of frames, since dimensionless observables will yield the exact same answer in any other conformal frame.

Let us provide another common example of a choice of frame. Suppose that the action in Eq. \eqref{eq:scalartensoraction} is supplemented with a matter Lagrangian $\mathcal{L}_m[g_{\mu\nu};\boldsymbol{\psi}]$, and that the interactions between matter fields $\boldsymbol{\psi}$ and non-minimally coupled scalars $\bphi$ are mediated purely through gravity. This presentation of the action is usually defined as the Jordan frame. By extracting the conformal factor $\Omega$ and keeping it explicit in the action, it is clear that there is also a continuum of Jordan frames
\begin{equation}\label{eq:Jordanframecondition}
	\mathsf{J}\colon\qquad \Omega^{d-2} = \lambda^{d-2}
\end{equation}
where $\lambda>0$ is a dimensionless constant (the exponential factor is purely for later convenience). Here it is perhaps more intuitive to see that $\{\mathsf{J}(\lambda)\}$ is just the ``horizontal'' foliation of $\mathcal{M}^\omega$. Moreover, we can now view frame transformations as leaf space bijections, since they take us between hypersurfaces belonging to different foliations. For instance, the transformation from Jordan to Einstein frame can be written as a map $\mathcal{F}\colon \mathcal{M}^\omega\to \mathcal{M}^\omega$ with
\begin{eqnarray}
	\mathcal{F}\colon & \mathsf{J}(\lambda) & \longmapsto \ \mathsf{E}(\lambda M_\mathsf{E})\,, \\
	& (\omega,\phi^i)=(\ln \lambda, \phi^i) & \longmapsto \left( \ln\lambda + \ln \frac{ M_\mathsf{E}}{f(\bphi)^\frac{1}{d-2} }, \phi^i\right)\,. \nonumber
\end{eqnarray}
Because the invariance of dimensionless ratios require frame- and unit transformations to be synchronous, the map $\mathcal{F}$ also keeps track of how systems of units must map between hypersurfaces. To see this in practice, consider some globally defined unit on $\mathcal{M}^\omega$; a very natural choice is the effective gravitational coupling $\mathrm{M}_\text{eff}(\bPhi) = \Omega f(\bphi)^\frac{1}{d-2}$. One can easily check when restricted to Einstein frame, $\imath^*_\mathsf{E}\mathrm{M}_\text{eff}=M_\mathsf{E}$. By taking the pullback along $\mathcal{F}^{-1}$, compare this with the corresponding unit on $\mathsf{J}(1)=\mathcal{F}^{-1}\mathsf{E}(M_\mathsf{E})$ that yields the same dimensionless ratios,
\begin{equation}
    ((\mathcal{F}^{-1})^* \imath^*_\mathsf{J}\mathrm{M}_\text{eff})(\bPhi) = f(\bphi)^\frac{1}{d-2}\,.
\end{equation}
In other words, working in Jordan frame and in units of $f(\boldsymbol{\phi})^\frac{1}{d-2}$ yields the \emph{exact same} dimensionless observables as working in Einstein frame and in units of $M_\mathsf{E}$.\footnote{In fact, Dicke already highlighted this result back in the 1960s \cite{Dicke:1961gz}, yet the frame problem somehow prevailed many years later.} These are two completely equivalent physical descriptions, \emph{provided} one makes the appropriate\footnote{By `appropriate change of units', we can now be precise and refer the pullback of the associated frame transformation. This applies in general: if $\mathrm{M}(\bPhi)$ is a unit on $\mathsf{\Sigma}$ and $\mathcal{F}\colon\mathsf{\Sigma}'\mapsto \mathsf{\Sigma}$ is a frame transformation, then $((\mathcal{F}^{-1})^*\mathrm{M})(\bPhi)$ is the compatible unit on $\mathsf{\Sigma}'$. Note that $\mathrm{M}\propto \Omega$ to be consistent with Eq. \eqref{unitchange}.} change of units.

\subsection{ADM Formalism in Field Space}
So far we have motivated a novel perspective of thinking about conformal frames as foliating hypersurfaces in augmented field space, and also described frame- and unit transformations in geometric terms. Our next aim is to develop the tools necessary to understand how quantities evolve under field excursions, which will be central for discussing the SDC and SSDC in Section \ref{sec:distanceconjecture}. 

We remind that (in the metric-first point of view), after explicitly extracting the conformal factor in the action and identifying the metric $\mathsf{G}_{IJ}$ on $\mathcal{M}^\omega$, we always have the freedom to make two independent choices: 
\begin{enumerate}
    \item A conformal frame $\mathsf{\Sigma}\subset \mathcal{M}^\omega$, which will tell us the induced field space metric $\mathsf{G}^\mathsf{\Sigma}_{ij}$.
    \item A unit $\mathrm{M}$ which we wish to measure dimensionful quantities in terms of.
\end{enumerate}
It is only when both are provided that it becomes possible to compute dimensionless quantities. More importantly, the two choices enable us to compute how quantities \emph{vary} with respect to other dimensionless parameters; for example, under field excursions where the scalar vevs act as coordinates on $\mathcal{M}^\omega$. Indeed, we already encountered the unit-covariant derivative $\mathrm{D}_\mu$, which, when provided a unit, measures the \emph{physical} spacetime variation of dimensionful quantities (that is to say their variation \emph{up to} the choice of unit). The objective for this section is likewise to formulate an analogue derivative on field space; one which is field-, unit-, \emph{and} frame-covariant.

The fact that a conformal frame $\mathsf{\Sigma}$ gives rise to an entire foliation of $\mathcal{M}^\omega$ implies that we can apply a very natural language to describe the field space geometry on $\mathsf{\Sigma}$: the ADM formalism. There is actually a striking similarity between how the ADM formalism is usually applied to decompose time and space, and how we will decompose the augmented field space, as $\omega$ corresponds to the ``timelike'' direction in $\mathcal{M}^\omega$. Every $\mathsf{\Sigma}$ of interest in this section will be a ``spacelike'' hypersurface. We will assume familiarity with the basics of ADM; for an introduction, we refer the interested reader to \cite{Corichi:1991qqo, Gourgoulhon:2012ffd}.

Starting with $\mathsf{G}_{IJ}$ as given in Eq.~\eqref{eq:augmentedmetricdimless}, we consider some frame $\mathsf{\Sigma}$ whose corresponding foliation of $\mathcal{M}^\omega$ is defined by the level sets of the function $\sigma(\bPhi)$. This allows us to calculate the normal vector to $\mathsf{\Sigma}$ through
\begin{equation}
    \mathsf{n}_I = \pm \frac{\partial_I\sigma}{\sqrt{\abs{\mathsf{G}^{IJ}(\partial_I\sigma)(\partial_J\sigma)}}}\,.
\end{equation}
Usually the sign is chosen such that $\mathsf{n}^I$ points in the direction of increasing $\omega$. In the two examples of Einstein and Jordan frame that we covered in the previous section, specified via Eqs. \eqref{eq:Einsteinframecondition} and \eqref{eq:Jordanframecondition}, one finds
\begin{align}
    \label{eq:Einsteinnormaldown}
    \mathsf{E}\colon \qquad & \mathsf{n}_I = -\sqrt{\frac{d-1}{d-2} \left(\frac{M_\mathsf{E}}{\mathrm{M}}\right)^{d-2}}\begin{pmatrix}
        d-2\\ \frac{f_{,i}}{f}
    \end{pmatrix}\,,\\
    \mathsf{J}\colon \qquad & \mathsf{n}_I = -\sqrt{(d-1)(d-2)\left(\frac{\lambda}{\mathrm{M}}\right)^{d-2} \left[f+\frac{d-1}{d-2}\mathsf{G}^{ij}f_{,i}f_{,j}\right]}\begin{pmatrix}
        1\\ 0
    \end{pmatrix}\,.
\end{align}
Likewise, one usually defines an evolution vector~$\mathsf{t}^I$, analogous to the time-evolution vector, which Lie drags $\mathsf{\Sigma}$ in the direction of increasing~$\omega$. Combined with the normal vector, these define the lapse and shift functions for the corresponding foliation of augmented field space. They are related by
\begin{equation}
    \mathsf{t}^I \equiv \frac{1}{\partial_\omega\sigma}\delta_\omega^I = \mathsf{N}\mathsf{n}^I + \mathsf{N}^I\,.
\end{equation}
Note that $[\mathsf{t}^I]=[\mathsf{N}]=[\mathsf{N}^I]=-[\sigma]$ can be dimensionful in our convention, and depends on how one parameterises the function $\sigma$. Here $\mathsf{N}^I$ is the shift vector, obeying $\mathsf{N}^I\mathsf{n}_I = 0$, and $\mathsf{N} = -\mathsf{t}^I\mathsf{n}_I$ is the lapse function. Using the same two examples, their respective lapse and shift turn out to be
\begin{align}
     \mathsf{E}\colon \quad & &\mathsf{N}^2 &= \frac{d-1}{d-2}(M_\mathsf{E}\mathrm{M})^{-(d-2)}\,, &  \mathsf{N}^I  &= 0 \,, \\
     \mathsf{J}\colon \quad & &\mathsf{N}^2 &= \frac{d-1}{d-2}(\lambda \mathrm{M})^{-(d-2)}\left[f+\frac{d-1}{d-2}\mathsf{G}^{ij}f_{,i}f_{,j}\right]\,,  &\mathsf{N}^I &= -\frac{d-1}{d-2}\lambda^{-(d-2)}\begin{pmatrix}
         0 \\ \mathsf{G}^{ij}f_{,j}
     \end{pmatrix}\,.
\end{align}
It is interesting to note that Einstein frame has vanishing shift. In fact, for a completely general implicit function $\sigma(\bPhi)$, the shift is
\begin{equation}
    \mathsf{N}^I \propto \sigma_{,i} - \frac{1}{d-2}\frac{f_{,i}}{f} \sigma_{,\omega}\,.
\end{equation}
So provided some simple assumption that $\sigma(\bPhi) = \sigma_1(\omega)\sigma_2(\bphi)$, Einstein frame turns out to be the \emph{unique} frame where the shift vanishes. In Section~\ref{sec:geodesics} we comment further on some nice  features of Einstein frame. Now, using the normal vector, one can also obtain the induced field space metric on $\mathsf{\Sigma}$ via the pullback of
\begin{equation}
    \mathsf{G}_{IJ}^\mathsf{\Sigma} = \mathsf{G}_{IJ} + \mathsf{n}_I \mathsf{n}_J\,.
\end{equation}
Indeed, we recover the results obtained in the previous section for the induced metric for the Einstein and the Jordan foliations,
\begin{align}
    \label{eq:Einsteininducedmetric}
    \mathsf{G}^\mathsf{E}_{ij} &= \left(\frac{M_\mathsf{E}}{\mathrm{M}}\right)^{d-2}\left(\frac{\widehat{\mathsf{G}}_{ij}}{f} + \frac{d-1}{d-2}\frac{f_{,i}f_{,j}}{f^2}\right)\,,\\
    \mathsf{G}^\mathsf{J}_{ij} &= \left(\frac{\lambda}{\mathrm{M}}\right)^{d-2} \widehat{\mathsf{G}}_{ij}\,.
\end{align}
Finally, one can address how some dimensionless quantity $\Pi$ changes along a particular direction in~$\mathsf{\Sigma}$ using the induced field-covariant derivative. Let $\nabla_I$ denote the covariant derivative associated to the augmented field space metric $\mathsf{G}_{IJ}$. The induced covariant derivative on $\mathsf{\Sigma}$ is
\begin{equation}
    \nabla_I^\mathsf{\Sigma}\Pi \equiv \tensor{\mathsf{G}}{^{\mathsf{\Sigma}J}_I} \, \nabla_J\Pi = \nabla_I\Pi + \mathsf{n}_I\mathsf{n}^J\nabla_J\Pi\,.
\end{equation}
This derivative indeed measures the \emph{physical} rate of change of $\Pi$ in the conformal frame $\mathsf{\Sigma}$. If $\Pi$ was dimensionful, the above expressions would not transform covariantly under unit transformations. Therefore, we must generalise even further and define a field-, frame-, \emph{and} unit-covariant derivative, by adding a term analogous to Eq.~\eqref{unitcovder}. Suppose $Q$ is a dimensionful quantity and $\mathrm{M}$ is a choice of unit. Then
\begin{equation}\label{eq:fullder}
    \mathrm{D}^\mathsf{\Sigma}_I Q \equiv \nabla_I^\mathsf{\Sigma}Q - [Q] \frac{\nabla_I^\mathsf{\Sigma}\mathrm{M}}{\mathrm{M}}Q
\end{equation}
measures the physical rate of change of $Q$ on $\mathsf{\Sigma}$.\footnote{\label{footnote:derivativetransformation}
The two derivatives are `covariant' in the standard sense: consider some $\Pi(\bPhi)$, $Q(\bPhi)$, and unit $\mathrm{M}(\bPhi)$ on $\mathsf{\Sigma}$, a tangent vector $\mathsf{v}\in T_{\bPhi}\mathsf{\Sigma}$, and a frame transformation $\mathcal{F}\colon \mathsf{\Sigma}'\mapsto \mathsf{\Sigma}$. Directional derivatives on $\mathsf{\Sigma}$ and $\mathsf{\Sigma}'$ are then related by
\begin{align}
    (\nabla_\mathsf{v}^\mathsf{\Sigma} \Pi)(\bPhi) &= \left(\nabla^{\mathsf{\Sigma}'}_{(\mathcal{F}^{-1})_*\mathsf{v}} (\mathcal{F}^*\Pi)\right)(\mathcal{F}^{-1}(\bPhi))\,,\\\
    (\DD_\mathsf{v}^\Sigma Q)(\bPhi)&= \left( \nabla^{\mathsf{\Sigma}'}_{(\mathcal{F}^{-1})_*\mathsf{v}}(\mathcal{F}^*Q) - [Q]\frac{\nabla^{\mathsf{\Sigma}'}_{(\mathcal{F}^{-1})_*\mathsf{v}}(\mathcal{F}^* \mathrm{M})}{\mathcal{F}^*\mathrm{M}}(\mathcal{F}^*Q)\right)(\mathcal{F}^{-1}(\bPhi))\,.
\end{align}
}
 
This fully covariant derivative features the same properties as the ``vanilla'' unit-covariant derivative presented in Section \ref{sec:unittransformations}. E.g. $\DD_I^\mathsf{\Sigma}Q = \mathrm{M}^{[Q]}\nabla_I^\mathsf{\Sigma}\Pi$. Furthermore, there exists a special choice of unit where the unit-covariant derivatives reduce to the ordinary covariant derivative. This ``preferred unit'' is the one which remains covariantly constant on the manifold of interest. For instance, on flat spacetime, the unit $\mathrm{M}=\text{const.}$ makes computation much simpler, because $\partial_\mu \mathrm{M}=0$ and hence $\DD_\mu Q=\nabla_\mu Q$. The same reasoning applies to augmented field space: a preferred unit is one which is covariantly constant on $\mathsf{\Sigma}$, which is to say that $\nabla_I^\mathsf{\Sigma}\mathrm{M}=0$. This system of units must be a function purely of the foliation parameter $\sigma$. Taking Einstein frame as an example, $\nabla_I^\mathsf{\Sigma}M_\mathsf{E}=0$, which makes $\mathrm{M}=M_\mathsf{E}$ a ``preferred unit'' in Einstein frame. Likewise, in Jordan frame, any $\mathrm{M} = \mathrm{M}(\omega)$ that is independent of $\bphi$ is a preferred unit.

\subsection{Geodesics and Distance}\label{sec:geodesics}
In the final part of this section, let us focus our attention on the meaning of field space geodesics and geodesic distance. This discussion is of course inspired by the Distance Conjectures, which are examined later in Section~\ref{sec:distanceconjecture}. For that purpose we are essentially interested in formulating a well-defined notion of distance in field space. Provided a conformal frame, an intuitive ansatz is 
\begin{equation}\label{eq:distance}
    \Delta_\phi^\mathsf{\Sigma} = \int_{\gamma\subset \mathsf{\Sigma}}\dd s\sqrt{\mathsf{G}^\mathsf{\Sigma}_{ij}(\bphi) \dv{\phi^i}{s}\dv{\phi^j}{s}}\,,
\end{equation}
where $\gamma$ is a geodesic with respect to the induced metric $\mathsf{G}_{ij}^\mathsf{\Sigma}$. However, it is immediately obvious that Eq.~\eqref{eq:distance} presents a dilemma when it comes to the choice of frame. For gravitational EFTs, there are an infinite number of frames $\mathsf{\Sigma}$ to choose from, and not all of them yield equally sensible induced metrics. Consider for example the following theory written respectively in Jordan and Einstein frame,
\begin{align}
    S^\mathsf{J} &= \int \dd^dx \sqrt{-g}\left[\frac{f(\bphi)}{2}R[g]\right]\,,\\
    S^\mathsf{E} &= \int \dd^dx \sqrt{-g}\left[\frac{M_\mathsf{E}^{d-2}}{2}R[g] - \frac{1}{2}M_\mathsf{E}^{d-2} \frac{d-1}{d-2}\frac{f_{,i}f_{,j}}{f^2}g^{\mu\nu}\partial_\mu\phi^i\partial_\nu\phi^j\right]\,.
\end{align}
If one was to directly apply Eq.~\eqref{eq:distance} across both frames, it would lead to contradicting results: on one hand, the Einstein frame produces generally finite positive distances, whereas $\mathsf{G}^\mathsf{J}_{ij} = 0$, and thus all distances vanish in Jordan frame. In general $\Delta_\phi^\mathsf{\Sigma}\neq \Delta_\phi^{\mathsf{\Sigma}'}$. 

\begin{figure}[H]
    \centering
    	\includegraphics[scale=0.50]{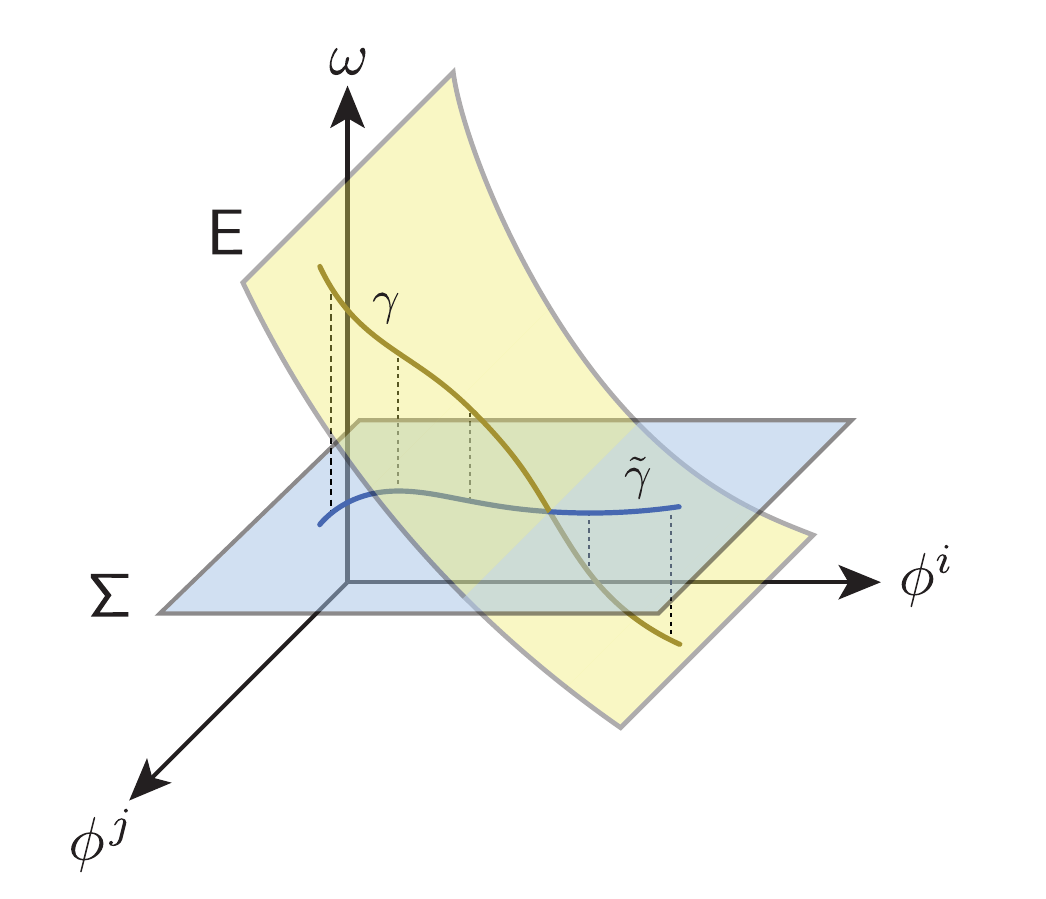}
    \caption{If $\gamma$ is a spacelike geodesic with respect to the augmented field space metric $\mathsf{G}_{IJ}$, then it will also be a geodesic on Einstein frame. Under a frame transformation $\gamma\mapsto \widetilde{\gamma}$ it is no longer guaranteed to be a geodesic.}
    \label{fig:geodesics}
\end{figure}

To disentangle the situation, we again look to the augmented field space $\mathcal{M}^\omega$. That is, consider some spacelike geodesic $\gamma\subset \mathcal{M}^\omega$ and let $\mathsf{v}\in \Gamma(T\mathcal{M}^\omega)$ be tangent to $\gamma$ at every $\bPhi\in \gamma$. Then $\mathsf{v}$ obeys the geodesic equation
\begin{equation}
    \nabla_\mathsf{v} \mathsf{v}^I = 0\,.
\end{equation}
We take $\mathsf{v}$ to be dimensionless such that we do not have to worry about the unit-covariant derivative. The associated geodesic distance in $\mathcal{M}^\omega$ is perfectly well-defined.
\begin{equation}
    \Delta_\Phi = \int_{\gamma\subset \mathcal{M}^\omega} \dd s \,\sqrt{\mathsf{G}_{IJ}(\bPhi)\dv{\Phi^I}{s}\dv{\Phi^J}{s}}
\end{equation}
Consider now any conformal frame $\mathsf{\Sigma}$. It is straightforward to show that
\begin{equation}
    \nabla_\mathsf{v}\mathsf{v}^I = \nabla_\mathsf{v}^\mathsf{\Sigma}\mathsf{v}^I - \mathsf{v}^J\mathsf{v}^K\mathsf{K}^\mathsf{\Sigma}_{JK}\mathsf{n}^I
\end{equation}
where $\mathsf{K}^\mathsf{\Sigma}_{IJ}$ is the extrinsic curvature of $\mathsf{\Sigma}$. Therefore, $\mathsf{\Sigma}$ is a \emph{totally geodesic hypersurface}, meaning that geodesics in $\mathcal{M}^\omega$ are also geodesics on $\mathsf{\Sigma}$, \emph{if and only if} the extrinsic curvature vanishes. Which frame does this correspond to? It is Einstein frame! By straightforward computation
\begin{equation}
    \mathsf{K}_{ij}^\mathsf{E} = -\frac{1}{2}\mathcal{L}_\mathsf{n}\mathsf{G}^\mathsf{E}_{ij} \propto (\mathrm{M}-\partial_\omega\mathrm{M})\,.
\end{equation}
Einstein frames thereby foliate augmented field space by totally geodesic hypersurfaces, provided $\mathrm{M}\propto \Omega$. This condition on the unit is already implied by the consistency with Eq.~\eqref{unitchange}, so $\mathsf{K}_{ij}^\mathsf{E}=0$. In any other frame the extrinsic curvature is generally non-vanishing. We may conclude that geodesics and geodesic distance in augmented field space precisely match their respective notions in Einstein frame,
\begin{equation}
    \Delta_\Phi = \Delta_\phi^\mathsf{E}\,.
\end{equation}
In summary, while we have argued that all conformal frames provide synonymous descriptions of the same physics, not all frames prove to be equally useful when it comes to studying field dynamics. Here we have seen that it is crucial to work in the Einstein frame if we want to recover the correct geodesic distance. Any other frame will return the wrong answer, simply because curves $\widetilde{\gamma}\subset \mathsf{\Sigma}\neq \mathsf{E}$ cannot be true geodesics. Rather, these trajectories are $\widetilde{\gamma}=\gamma \slash{\sim}$, where $\gamma$ is the geodesic in $\mathcal{M}^\omega$, obtained by quotienting or `projecting out' frame transformations (see Figure \ref{fig:geodesics}). In doing so, we automatically lose information about the dynamics as well as distance traversed along the normal of $\mathsf{\Sigma}$. The only foliation where $\gamma$ is contained in a single leaf is Einstein.

\section{From Frame Covariance to the Swampland Distance Conjecture}
\label{sec:distanceconjecture}

So far, we have outlined the fundamentals of frame covariance and presented our framework for the frame-augmented field space. In this section, we will apply these concepts to a setting native to the Swampland Programme, with the objective to appraise the SDC and SSDC in light of frame covariance.

We shall restrict our focus to diagonal toroidal compactifications, which both capture a large class of instructive models as well as provide a basic template for more involved flux compactifications in string theory. Perhaps the most famous examples where this kind of background is imposed are Kaluza-Klein theory and Freund-Rubin type flux compactifications, which are often covered as pedagogical examples to demonstrate several of the Swampland Conjectures \cite{VanRiet:2023pnx}. For example, it is well-known that these models accommodate infinite towers of KK modes whose scaling behaviour complies both with the SDC and SSDC. We therefore take these models as our starting point for similar pedagogical reasons and to set the scene for future analyses.

To stay as general as possible, we consider some rather arbitrary warped product spacetime manifold ${\mathfrak{M}_D=\mathfrak{M}_d\times_w\mathfrak{X}_1\times_w\cdots \times_w \mathfrak{X}_N}$. Here $\mathfrak{M}_d$ is $d$-dimensional spacetime and $\mathfrak{X}_i$ are compact manifolds of dimension $n_i$, where $D=d+\sum_i n_i$. Sometimes these are referred to as the \emph{external} and \emph{internal} parts of $\mathfrak{M}_D$, respectively. The objective when compactifying the theory from $D$ to $d$ dimensions is essentially to integrate the action over the internal manifolds $\mathfrak{X}_1,\dots,\mathfrak{X}_N$, which results in a $d$-dimensional EFT on $\mathfrak{M}_d$. 

We begin with an ansatz for the warped product metric
\begin{equation}\label{eq:compactificationansatz}
	\dd s^2 = G_{MN}\dd X^M\dd X^N = e^{2\omega}\dd s_{\mathfrak{M}_d}^2 + \sum_{i=1}^N e^{2\phi^i}\dd s_{\mathfrak{X}_i}^2\,.
\end{equation}
Here, the moduli $\Phi^I = (\omega,\phi^i)$ are taken to be functions of \emph{external} spacetime coordinates only. Note that we use the same labels for the moduli $\Phi^I$ as for the augmented field space coordinates. This is no coincidence. After compactifying to $d$ dimensions, the moduli $\Phi^I$ manifest as scalar fields on $\mathfrak{M}_d$ in the effective action. The $d$-dimensional dilaton $\omega$ thus essentially acts as the conformal mode: its field excursions effectively induce Weyl transformations on the external part of the bulk metric. We can therefore identify the moduli space with the augmented field space $\mathcal{M}^\omega$.

Consider then pure $D$-dimensional gravity, described by the Einstein-Hilbert action
\begin{equation}
	S = \int \dd^DX \sqrt{-G}\frac{\MplD^{D-2}}{2}R[G]\,.
\end{equation}
As per usual, $\MplD$ is the $D$-dimensional Planck mass, with $[\MplD]=1$. Note that the Einstein-Hilbert action is by definition in Einstein frame, since $\MplD$ is independent of any fields. By careful computation, one can show that the compactified theory takes on the form
\begin{equation}\label{eq:compactifiedaction}
	S = \int \dd^dx\sqrt{-g} \left[\frac{\MplD^{D-2}\mathcal{V}_\mathfrak{X}}{2} e^{(D-2)\varpi(\bPhi)} R[g] - \frac{1}{2} g^{\mu\nu} \widehat{\mathsf{G}}_{IJ}(\bPhi)\partial_\mu\Phi^I \partial_\nu\Phi^J + \cdots \right]\,.
\end{equation}
Here $g_{\mu\nu}$ is the metric on $\mathfrak{M}_d$ and $R[g]$ is the associated Ricci scalar curvature. Ellipses denote the omitted scalar curvature terms of the internal manifold(s), which will contribute to an effective potential for the moduli. These will not be relevant for this discussion, as we do not deal with the question of moduli stabilisation. Next, the (dimensionful) constant $\mathcal{V}_\mathfrak{X}$ is defined as
\begin{equation}
\mathcal{V}_\mathfrak{X} = \int_{\mathfrak{X}_1}\dd^{n_1}y_1\sqrt{g_{\mathfrak{X}_1}}\cdots \int_{\mathfrak{X}_N}\dd^{n_N}y_N \sqrt{g_{\mathfrak{X}_N}},
\end{equation}
with dimension $[\mathcal{V}_\mathfrak{X}]=-(D-d)$. Note that this constant is distinct from the typically defined moduli-dependent ``physical volume'' or ``warped volume'' of the internal manifold
\begin{equation}\label{eq:physicalvolume}
\begin{split}
	\mathcal{V}_w(\bphi) &\equiv \int_{\mathfrak{X}_1}\dd^{n_1}y_1\sqrt{G_{\mathfrak{X}_1}}\cdots \int_{\mathfrak{X}_N}\dd^{n_N}y_N \sqrt{G_{\mathfrak{X}_N}}\,,\\
    &=\mathcal{V}_\mathfrak{X}e^{n_i\phi^i}\,.
\end{split}
\end{equation}
In the study of compactifications, the warped volume is usually used to \emph{define} the effective $d$-dimensional Planck mass through the relation
\begin{equation}\label{eq:Mpld}
	\Mpld^{d-2}(\bphi)\equiv \MplD^{D-2}\mathcal{V}_w(\bphi)\,.
\end{equation}
In our notation, the exponential factor in Eq.~\eqref{eq:physicalvolume} appears in the relationship through the function
\begin{equation}
	(D-2)\varpi(\bPhi) \equiv (d-2)\omega + n_i \phi^i
\end{equation}
giving the well-known functional dependence of $\mathcal{V}_w(\bphi)$ in terms of the radions. With some foresight we have included a factor $(D-2)$ in the definition of $\varpi$, which we return to in Section~\ref{sec:compactificationEinsteinframe}. Note the familiar factor $e^{(d-2)\omega}$ coming from the $d$-dimensional Weyl transformations.

It is prudent to think of $\mathcal{V}_\mathfrak{X}$ as depending only on the choice of coordinates on the internal manifold. A simple example is when $\mathfrak{X} = S^1$: we can readily see that different choices of coordinate, e.g. $[0,1)$ or $[0,2\pi)$, will yield different volume factors once integrated. However, we acknowledge such difference to have no physical bearing and is just an artefact of our parameterisation. Indeed, we have already argued that such a choice can be scaled away by a corresponding constant frame transformation. Therefore, we can choose ${\mathcal{V}_\mathfrak{X}=\MplD^{-(D-d)}}$ without loss of generality. This also implies that ${\Mpld> \MplD}$ for super-Planckian volumes. That is, the true gravitational strong coupling scale is generically expected to be below $\Mpld$. Nonetheless, we keep $\mathcal{V}_\mathfrak{X}$ explicit in order to better keep track of dimensionful quantities that appear in our analysis. 

Finally, one can show straightforwardly by induction that the dimensionless augmented field space metric reads
\begin{equation}\label{eq:compactificationaugmentedmetric}
	\mathsf{G}_{IJ}(\bPhi) = \frac{1}{\mathrm{M}^{d-2}}\widehat{\mathsf{G}}_{IJ}(\bPhi)=-\frac{\MplD^{D-2}\mathcal{V}_\mathfrak{X}}{\mathrm{M}^{d-2}} e^{(D-2)\varpi(\bPhi)} \left(\mathsf{d}_I\mathsf{d}_J-\mathsf{d}_{IJ}\right)
\end{equation}
where
\begin{equation}
	\mathsf{d}_I = \begin{pmatrix}
		d-1\\
		n_1\\
		\vdots \\
		n_N
	\end{pmatrix}_I\,,\qquad \mathsf{d}_{IJ}=\begin{pmatrix}
		d-1 & 0 & \cdots & 0\\
		0 & n_1 & & \vdots\\
		\vdots & & \ddots & 0 \\
		0 & \cdots & 0 & n_N
	\end{pmatrix}_{IJ}\,.
\end{equation}
We therefore arrive at the exact same picture described in Section \ref{sec:augmentedfieldspace}: hypersurfaces in $\mathcal{M}^\omega$ correspond to different choices of conformal frames whose geometry is induced from $\mathcal{M}^\omega$.

\subsection{On Planck Masses and Einstein Frame}\label{sec:compactificationEinsteinframe}
Now that we have established the augmented field space geometry, the next step is to target Einstein frame, which will be a totally geodesic hypersurface. To do so, we follow the formalism laid out in previous sections. Recall that Einstein frame is the uniquely defined frame (up to constant Weyl transformations) in which the scalars are minimally coupled to gravity. From Eq.~\eqref{eq:compactifiedaction}, it is evident that Einstein frame is defined by the level sets of $\varpi(\bPhi)$, and that the field-\emph{in}dependent effective Planck mass is given by
\begin{equation}
\begin{split}
    \mathsf{E}\colon \quad \varpi(\bPhi) &= \text{const.,}\\
    M_\mathsf{E}^{d-2} &= \MplD^{D-2}\mathcal{V}_\mathfrak{X}e^{(D-2)\varpi}\,.
\end{split}
\end{equation}
If $\mathcal{V}_\mathfrak{X}\propto \MplD^{-(D-d)}$, then we find that $M_\mathsf{E}\propto \MplD$, and as such it is directly related to the gravitational cutoff in $D$ dimensions. We then compute the normal vector
\begin{equation}
    \mathsf{n}_I = -\sqrt{\frac{d-1}{d-2}\left(\frac{M_\mathsf{E}}{\mathrm{M}}\right)^{d-2}}\begin{pmatrix}
        d-2\\ n_i
    \end{pmatrix}\,,
\end{equation}
as well as the induced metric
\begin{equation}
    \mathsf{G}_{ij}^\mathsf{E}(\bphi) = \left(\frac{M_\mathsf{E}}{\mathrm{M}}\right)^{d-2} \left(\frac{1}{d-2}\mathsf{d}_i\mathsf{d}_j + \mathsf{d}_{ij}\right)\,.
\end{equation}
Notice again the similarities with the general scalar-tensor results in Eqs. \eqref{eq:Einsteinnormaldown} and \eqref{eq:Einsteininducedmetric}. Taking the base unit to coincide with the Einstein frame Planck mass $\mathrm{M}=M_\mathsf{E}$, therefore giving us $\nabla_I^\mathsf{E}M_\mathsf{E}=0$, we can further verify the vanishing of the extrinsic curvature $\mathsf{K}_{ij}^\mathsf{E}=0$.

The above example makes it clear that Einstein frame, as defined in scalar-tensor theory, is different than the Einstein frame commonly referred to in the string compactification literature. In the context of compactifications, Einstein frame usually refers to the frame where $\Mpld^{d-2}=\MplD^{D-2}\mathcal{V}_w(\bphi)$, as defined in Eq.~\eqref{eq:Mpld}, appears in front of the Ricci scalar. However, $\Mpld(\bphi)$ is sensitive to perturbation theory thanks to its dependence on the fields, and thus $\bphi$ couples non-minimally to gravity. Therefore, this definition is in direct contrast to the perspective of scalar-tensor theory, which stipulates that, in Einstein frame, there should be no field dependence baked into the prefactor of the Ricci scalar. As a result, the ``$d$-dimensional Planck mass'', $\Mpld$, and the Einstein frame effective Planck mass, $M_\mathsf{E}$, are mismatched by a crucial conformal factor
\begin{equation}\label{eq:MEcompactification}
    M_\mathsf{E}^{d-2} = e^{(d-2)\omega}\Mpld^{d-2}(\bphi)\,. 
\end{equation}
Compare this with the definition of the effective Planck mass given in Eq.~\eqref{eq:Einsteinframecondition}. Therefore, it is more natural to think of $\Mpld$ as corresponding to the non-minimal coupling $f(\bphi)^\frac{1}{d-2}$ in the Jordan frame.\footnote{We could effectively omit $\omega$ by fixing it at any constant value, which would make it seem that $\Mpld(\bphi)$ is the effective $d$-dimensional Planck mass. However, $\omega = \text{const.}$ precisely defines the Jordan frame as seen in Eq. \eqref{eq:Jordanframecondition}, whose effective Planck mass reads $f(\bphi)^\frac{1}{d-2}$.} Equivalently, one can show that the non-minimal Jordan frame coupling pulled back to Einstein frame $\mathrm{M}_\mathsf{J}(\bphi)=(\mathcal{F}^{-1})^*\imath^*_\mathsf{J}\mathrm{M}_\text{eff}(\bPhi)$ is consistent with the definition of $\Mpld$.

Another way to demonstrate that $\varpi = \text{const.}$ is an equivalent specification of Einstein frame is by applying the following redefinitions $(\omega,\bphi)\mapsto (\varpi,\brho)$ to the original metric ansatz in $D$ dimensions:
\begin{equation}\label{eq:Ddimensionalmodulibasis}
\begin{split}
	\omega = \varpi &+ \rho_0\,,\qquad \phi^i = \varpi + \rho^i\,,\qquad \rho_0\equiv -\frac{n_i}{d-2}\rho^i\,,\\
	&\dd s^2 = e^{2\varpi}\left(e^{2\rho_0} \dd s_{\mathfrak{M}_d}^2 + \sum_{i=1}^N e^{2\rho^i} \dd s^2_{\mathfrak{X}_i}\right)\,.
\end{split}
\end{equation}
It should be clear that $\varpi$ is simply the $D$-dimensional dilaton, whose field excursions amount to $D$-dimensional Weyl transformations.\footnote{For this reason, $(\omega,\bphi)$ are sometimes referred to as the ``$d$-dimensional dilaton/radions'', whereas $(\varpi,\brho)$ are the ``$D$-dimensional dilaton/radions'', see for example Appendix B of \cite{Shiu:2023fhb}.} Substituting this ansatz into the Einstein-Hilbert action, we indeed recover the correct conformal factor $e^{(D-2)\varpi}$ that we expect to appear in $D$ dimensions. Therefore, any Einstein frame can be obtained by a constant $D$-dimensional Weyl transformation imposing the condition $\varpi= \text{const.}$, giving us $M_\mathsf{E}\propto \MplD$.

\begin{figure}[ht!]
    \centering
    	\includegraphics[scale=0.5]{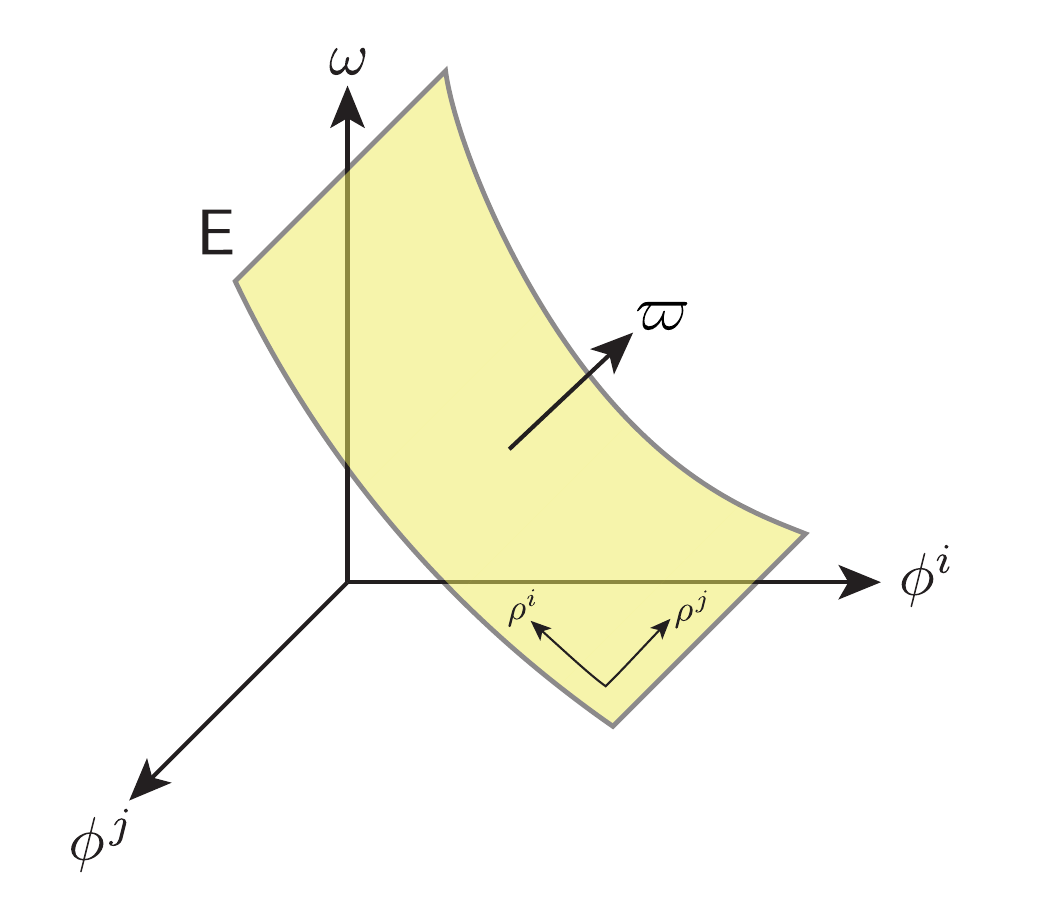}
    \caption{Sketching the $d$-dimensional moduli basis $(\omega,\bphi)$ and $D$-dimensional moduli basis $(\varpi,\brho)$ in the frame-augmented field space. For an inherently higher-dimensional theory originally in Einstein frame before compactification, the timelike direction is $\varpi$ rather than $\omega$.}
    \label{fig:coordinates}
\end{figure}

Without loss of generality, we may pick the Einstein frame at $\varpi = 0$, essentially omitting $\varpi$. Those familiar with string theory compactifications will recognize the resulting line element in Eq.~\eqref{eq:Ddimensionalmodulibasis} as an incredibly common metric ansatz employed in the literature for compactifications (see for instance \cite{VanRiet:2023pnx,Etheredge:2022opl,Etheredge:2024tok}). From the perspective of frame covariance, however, this ansatz corresponds to a particularly significant parameterisation. It turns out that $\{e^{2\rho_0},e^{2\rho^1},\dots,e^{2\rho^N}\}$, with $\rho_0$ defined as above, is the \emph{unique} choice of Weyl factors (up to field redefinitions) that guarantees radions $\brho$ to automatically be minimally coupled in the effective theory. It essentially ensures that the radions are adapted coordinates to Einstein frame; basis one-forms $\dd \rho^i$ on $\mathcal{M}^\omega$ restrict trivially to basis one-forms on $\mathsf{E}\subset \mathcal{M}^\omega$. Another way to put it is
\begin{center}
	Einstein frame in $D$ dimensions $\overset{\eqref{eq:Ddimensionalmodulibasis}}{\implies}$ Einstein frame in $d$ dimensions.
\end{center}
The distinction between the two sets of fields is illustrated in Figure \ref{fig:coordinates}. While the $(\omega,\bphi)$ basis provides intuition as to how field excursions correspond to Weyl transformations, the $(\varpi,\brho)$ basis inherently accommodates the Einstein foliation. 

In Section~\ref{sec:augmentedfieldspace}, we demonstrated how the conformal mode $\omega$ corresponds to a timelike direction in augmented field space. It is therefore tempting to think that the $d$-dimensional dilaton also yields a timelike direction in the geometry laid out in this section.  There are, however, two reasons why this cannot be the case. First, $d$-dimensional Weyl transformations do not scale the entire bulk geometry homogeneously, and therefore anything in the EFT sensitive to the relative size of external/internal geometry (such as Kaluza-Klein towers, see Section~\ref{sec:KKtowers}) will see $\omega$-excursions as a physically meaningful effect. Second, although the EFT is expressed in $d$ dimensions, the theory itself was always inherently $D$-dimensional to begin with and \emph{not} $d$-dimensional. It is therefore only expected to transform frame-covariantly under $D$-dimensional Weyl transformations. Indeed, we can determine the timelike direction by expressing the augmented field space metric in the $(\varpi,\brho)$ basis:
\begin{equation}
	\mathsf{G}_{AB}(\varpi) = \frac{\MplD^{D-2}\mathcal{V}_\mathfrak{X}}{\mathrm{M}^{d-2}}e^{(D-2)\varpi}
	\begin{pmatrix}
		-(D-1)(D-2) & \frac{D-2}{d-2}n_j\\
		\frac{D-2}{d-2}n_i & \mathsf{d}_{ij} + \frac{1}{d-2}\mathsf{d}_i\mathsf{d}_j
	\end{pmatrix},
\end{equation}
with indices $A,B\in (\varpi,\brho)$. It is then easy to check and confirm that $\varpi$ is the timelike direction. 

The fact that the timelike direction in $\mathcal{M}^\omega$ does not align with $\omega$ is a general feature of gravitational EFTs obtained via dimensional reduction. For example, the two dilatons $\omega$ and $\varpi$, related by Eq.~\eqref{eq:Ddimensionalmodulibasis}, will always point in different directions in $\mathcal{M}^\omega$. This is not without consequences, as we will see in the following discussions of the SSDC and SDC.

\subsection{The Species Scale Distance Conjecture}
As outlined in Section~\ref{sec:swampland}, an object of central importance in the Swampland Programme and in particular for the investigation of string compactifications is the species vector
\begin{equation}
    \mathcal{Z} \equiv -\nabla_i\ln\frac{\Lsp}{\Mpld}\,.
\end{equation}
Here $\nabla$ is the covariant derivative in moduli space. The species vector measures the rate of change of $\Lsp$ in units of $\Mpld$. Now, for theories such as the one expressed in Eq.~\eqref{eq:compactifiedaction}, it is expected that the species scale aligns with the gravitational cutoff of the original, non-compactified theory; i.e. the $D$-dimensional Planck scale $\MplD$. In other words, put in the language of frame covariance and scalar-tensor theory, where we have argued that $\MplD\leftrightarrow M_\mathsf{E}$ and $\Mpld \leftrightarrow M_\mathsf{J}$, the species vector is essentially given by
\begin{equation}\label{eq:speciesvectorcovariant}
    \mathcal{Z}_i \equiv \frac{1}{M_\mathsf{J}}\mathsf{D}_i^\mathsf{E} M_\mathsf{J} = -\nabla_i^\mathsf{E}\ln \frac{M_\mathsf{E}}{M_\mathsf{J}}\,.
\end{equation}
Thus, the vector in this scenario essentially measures the physical rate of change of the Jordan frame effective Planck mass in the Einstein frame. The second equality is expressed explicitly in Planck units, $\mathrm{M}=M_\mathsf{E}$. The virtue, however, of defining the species vector in terms of the frame- and unit-covariant derivative $\mathsf{D}$ is twofold. First, it is clear how the vector transforms when going to other conformal frames (see footnote \ref{footnote:derivativetransformation}). Second, our definition of $\mathcal{Z}$ needs no specification of units whatsoever; it is manifestly unit-independent! One could work in any other frame or unit and recover the exact same physically significant properties of $\mathcal{Z}$. 

The frame-covariant species vector defined in Eq.~\eqref{eq:speciesvectorcovariant} can be used to derive several key results. For instance, using the expressions for the field space metric and the Planck mass in Eqs.~\eqref{eq:Mpld}-\eqref{eq:MEcompactification}, we recover
\begin{equation}\label{eq:speciesvectornorm}
    \abs{\mathcal{Z}}^2 = \mathsf{G}^{\mathsf{E}\,ij} \mathcal{Z}_i\mathcal{Z}_j = \frac{1}{d-2}-\frac{1}{D-2}\,.
\end{equation}
This expression has been found to be a general property of the species vector when there are no other light towers except the leading KK tower \cite{Etheredge:2024tok}. However, in obtaining this result we have made no mention of KK towers whatsoever. In our language, the expression for the norm of the species vector given in Eq.~\eqref{eq:speciesvectornorm} follows as a result of the behaviour of the Jordan frame non-minimal coupling in Einstein frame. This appears to suggest that certain qualities of the species vector, as well as aspects of the SSDC, can be captured solely by the idea of frame covariance, independent of details of some underlying theory of quantum gravity. 

We recall that the Jordan frame, the way we have routinely referred to it, corresponds to hypersurfaces with $\omega = \text{const.}$, where $\omega$ is the $d$-dimensional dilaton. In ordinary $d$-dimensional scalar-tensor theories, Jordan frame is a spacelike hypersurface, as it corresponds to the horizontal foliation with respect to the timelike direction in $\mathcal{M}^\omega$. For inherently higher-dimensional theories, this is however not necessarily the case, since the timelike direction points along $\varpi$ different from $\omega$. For Jordan frame, Eq.~\eqref{eq:compactificationaugmentedmetric} reveals
\begin{equation}
    \mathsf{G}^{IJ}\mathsf{n}_I\mathsf{n}_J > 0 \quad\implies \quad \text{$\mathsf{J}$ is timelike.}
\end{equation}
Now, assuming that $\mathsf{G}_{ij}^\mathsf{J}$ is non-singular such that its inverse is well-defined, we can use the metric \eqref{eq:Einsteininducedmetric} to calculate the species vector given in Eq.~\eqref{eq:speciesvectorcovariant}. One finds
\begin{equation}\label{eq:speciesvectornormAB}
    \abs{\mathcal{Z}} = \frac{1}{\sqrt{(d-1)(d-2)}}\abs{\frac{\mathsf{B}}{\mathsf{A}}}
\end{equation}
where $\mathsf{A}$ and $\mathsf{B}$ arise from decomposing the Einstein frame normal 
\begin{equation}\label{eq:Einsteinnormalprojection}
    \mathsf{n}^I_\mathsf{E} = \mathsf{A}\mathsf{n}_\mathsf{J}^I + \mathsf{B}\mathsf{n}_\perp^I
\end{equation} 
in terms of the Jordan frame normal and another vector $\mathsf{n}_\perp$ perpendicular to $\mathsf{n}_\mathsf{J}$. More precisely, the coefficients are
\begin{align}
    \mathsf{A} &= \sqrt{\frac{d-1}{d-2}} \sigma_\mathsf{J} \mathsf{n}_\mathsf{J}^I\left(sf^{-1}\partial_If - \sigma_\mathsf{J}^{-1}\partial_I\sigma_\mathsf{J}\right)\,,\\
    \mathsf{B} &= \sqrt{\frac{d-1}{d-2}}\sigma_\mathsf{J} \abs{(\delta^I_K+s\mathsf{n}_\mathsf{J}^I\mathsf{n}_{\mathsf{J}\,K})f^{-1}\partial_I f}\,.
\end{align}
Here $s\equiv \mathrm{sign}(\mathsf{n}_\mathsf{J}^2)$ such that $s=+1$ for $\mathsf{J}$ timelike and $s=-1$ for $\mathsf{J}$ spacelike. One can intuitively think of Eq.~\eqref{eq:Einsteinnormalprojection} in analogy to boosting a timelike vector in special relativity. Consider two Lorentzian orthonormal bases $(e_0,e_1)$ and $(e'_0,e_1')$ related by a boost, with $e_0,e_0'$ timelike and $e_1,e_1'$ spacelike. In particular
\begin{equation}
    e_0 = \cosh(w)e_0' + \sinh(w)e_1'
\end{equation}
with $w$ being the rapidity. Taking the norm squared one straightforwardly confirms ${-1=-\cosh^2(w)+\sinh^2(w)}$. The two-dimensional slice spanned by $(\mathsf{n}_\mathsf{J},\mathsf{n}_\perp)$ is synonymous: we have a timelike unit vector $\mathsf{n}_\mathsf{E}$ recast in a basis of orthonormal time- and spacelike vectors, such that the coefficients $\mathsf{A},\mathsf{B}$ may be written in terms of the rapidity of this basis transformation. It leads to two cases depending on the `causal' nature of the Jordan frame.
\begin{equation}\label{eq:ABcases}
    \begin{cases}
        \mathsf{A}=\cosh w\,,\quad \mathsf{B} = \sinh w\, & \text{when $\mathsf{J}$ timelike ($s=+1$)}\\
        \mathsf{A}=\sinh w\,,\quad \mathsf{B} = \cosh w\, & \text{when $\mathsf{J}$ spacelike ($s=-1$)}
    \end{cases}
\end{equation}
That Jordan frame can be either space- or timelike is not a bizarre observation. The relation between the Einstein and Jordan frame metrics in Eq.~\eqref{eq:Einsteininducedmetric} is very reminiscent of a disformal transformation $g_{\mu\nu}\mapsto \mathcal{A}(\varphi) g_{\mu\nu} + \mathcal{B}(\varphi) \nabla_\mu\varphi \nabla_\nu\varphi$, which famously tilts the lightcone structure; following a disformal transformation, the norm of any null vector $v$ becomes $\mathcal{B}(v\cdot \nabla\varphi)^2$, with sign depending on $\mathcal{B}$. Albeit, when $\omega$ is the timelike direction, Jordan frame is guaranteed spacelike. That is the case for ordinary $d$-dimensional scalar-tensor theories. On the other hand, for diagonal toroidal compactifications (and we suspect also other models obtained from dimensional reduction), the timelike direction differs from $\omega$, and Jordan frame is allowed to become timelike. Different cases lead to different bounds for the species vector. Combining Eqs.~\eqref{eq:speciesvectornormAB} and \eqref{eq:ABcases} we summarise our result as follows:

\mbox{
\begin{tcolorbox}[boxrule=0pt,
    boxsep=0pt,
    colback={White!90!Gray},
    enhanced jigsaw,
    borderline west={2pt}{0pt}{Gray},
    before skip=10pt,
    after skip=10pt,
    sharp corners,
    breakable]
    For \emph{any} $d$-dimensional scalar-tensor theory that can be cast in a frame-covariant manner, the analogous species vector $\mathcal{Z}=M_\mathsf{J}^{-1}\mathsf{D}^\mathsf{E}M_\mathsf{J}$ obeys 
    \begin{equation}\label{eq:speciesscaleclaim}
        \abs{\mathcal{Z}} \begin{cases}
            \displaystyle <\frac{1}{\sqrt{(d-1)(d-2)}}\,, & \text{$\mathsf{J}$ spacelike}\\
            \displaystyle =\frac{1}{\sqrt{(d-1)(d-2)}}\,, & \text{$\mathsf{J}$ null}\\
            \displaystyle >\frac{1}{\sqrt{(d-1)(d-2)}}\,, & \text{$\mathsf{J}$ timelike}\\
        \end{cases}
    \end{equation}
\end{tcolorbox}
}
 
In the non-spacelike cases, we thus recover the lower bound for the SSDC. In Appendix \ref{sec:appendix} we provide further examples of models to demonstrate this result. One way to interpret Eq.~\eqref{eq:speciesscaleclaim} is essentially as a consistency criterion: if a scalar-tensor theory violates the appropriate bounds, it cannot be cast in a frame-covariant manner. It applies to a much broader class of scalar-tensor theories beyond string theory, suggesting that the SSDC has less to do with quantum gravity or UV completion \emph{per se}, and has more to do with how gravitational EFTs behave under Weyl transformations.\footnote{That is unless one has a reason to believe quantum gravity EFTs \emph{must} fit within a frame-covariant formalism. In that case, Eq.~\eqref{eq:speciesscaleclaim} qualifies as a Swampland constraint.} It is nevertheless interesting to ask whether compactifications or string theory EFTs generically admit non-spacelike Jordan frame hypersurfaces.

\subsection{The Sharpened Distance Conjecture}\label{sec:KKtowers}
Finally, we turn our attention to the SDC. Similar to the previous section, we will adopt a frame-covariant perspective which will allow us to recover a well-known bound. Since the SDC is related to towers of states in the effective theory, we begin by supplementing the Einstein-Hilbert action with a massless bulk scalar field $\chi$.
\begin{equation}
    S = S_\text{EH,$D$} + S_\chi= \int \dd^DX \sqrt{-G} \frac{\MplD^{D-2}}{2}\bigg[R[G] - G^{MN}\partial_M\chi\partial_N\chi \bigg]
\end{equation}
Using the ansatz of Eq.~\eqref{eq:compactificationansatz} (or equivalently Eq.~\eqref{eq:Ddimensionalmodulibasis} via a field redefinition), we find the following effective action after carefully integrating over the internal manifolds.
\begin{equation}
\begin{split}
    S &= S_\text{EH,$d$} + S_{\bPhi} + S_\text{KK}\\
    &= \int\dd^d x \sqrt{-g}\bigg[\frac{\MplD^{D-2}\mathcal{V}}{2}e^{(D-2)\varpi} R[g] - \frac{1}{2} g^{\mu\nu} \widehat{\mathsf{G}}_{IJ}(\bPhi)\partial_\mu \Phi^I \partial_\nu \Phi^J + \cdots \bigg]\\
    &\quad + \int \dd^dx \sqrt{-g} \MplD^{D-2}\mathcal{V}e^{(D-2)\varpi} \sum_{\vec{k}} \bigg[ -g^{\mu\nu}\partial_\mu \chi_{\vec{k}}^\dagger \partial_\nu \chi_{\vec{k}} + m_{\vec{k}}^2 \vert \chi_{\vec{k}}\vert^2 \bigg].
\end{split}
\end{equation}
To obtain this effective action, we have expanded $\chi$ into eigenmodes of the d'Alembertian of the internal space, whose associated eigenvalues $\lambda_{\vec{k}}$ are labelled by some vector $\vec{k}$. The KK modes $\chi_{\vec{k}}$ then obtain masses given by
\begin{equation}\label{eq:KKmass}
    m_{\vec{k}}^2(\bPhi) \simeq \mathrm{M}^2\sum_i \lambda_{k^i} e^{2\omega - 2\phi^i} = \mathrm{M}^2\sum_i \lambda_{k^i} e^{2\rho_0 - 2\rho^i}\,.
\end{equation}
The dependence on $\bPhi$ is essentially inherited from whatever Weyl rescaling factors contribute to $G_{\mu\nu}G^{mn}$. As per usual, $\mathrm{M}$ is the unit used in measuring the bulk line element $\dd s^2$. In a particular limit where some $\mathfrak{X}_i$ `becomes large', the tower which becomes light at the fastest rate goes as $m_\text{KK}\sim e^{\rho_0-\rho^i}$.

In simple terms, KK towers and their associated KK scale probe the `relative size' between internal and external manifold (cf. the $G_{\mu\nu}G^{mn}$ contribution to $m_\text{KK}$) and will be sensitive to any lower-dimensional Weyl transformation. In the language of moduli, the KK scale is sensitive to field excursions in the $\omega$ and $\phi^i$ directions. By the same argument, $m_\text{KK}$ must be invariant under $\varpi$ excursions  that induce $D$-dimensional Weyl transformations, since it preserves the relative scale of internal and external geometry. This is to say that $m_\text{KK}=m_\text{KK}(\rho)$ must be independent of $\varpi$, and is a function of Einstein frame coordinates only. 

Just as we did for the species vector, one can justify a definition of the charge-to-mass ratio or tower vector that is manifestly unit-independent
\begin{equation}
    \zeta \equiv -\frac{1}{m_\text{KK}}\mathsf{D^E}m_\text{KK}\,.
\end{equation}
We know that $m_\text{KK}$ behaves like a mass scale at the level of the effective theory in $d$ dimensions. From frame-covariant considerations, $m\propto e^\omega$ for any quantity with unit of mass, also in agreement with Eq.~\eqref{eq:KKmass}. The dependence of $m_\text{KK}$ as a function of $(\omega,\bphi)$ is therefore essentially fixed by the criterion that it must be independent of $\varpi$. Moreover, if the Einstein frame differs from $\omega = \text{const.}$, then $\dd\omega$ projected onto the Einstein frame will generally have a non-vanishing component along the direction of $\zeta$. By applying the Cauchy-Schwarz inequality, we therefore arrive at the following result:
\begin{tcolorbox}[boxrule=0pt,
    boxsep=0pt,
    colback={White!90!Gray},
    enhanced jigsaw,
    borderline west={2pt}{0pt}{Gray},
    before skip=10pt,
    after skip=10pt,
    sharp corners,
    breakable]
    For \emph{any} $d$-dimensional scalar-tensor theory containing some mass scale $m$ that can be cast in a frame-covariant manner such that $m\propto e^\omega$, the following inequality holds.
    \begin{equation}\label{eq:sharpenedbound}
        \abs{\zeta} \geq \frac{\langle \nabla^\mathsf{E}\omega,\zeta\rangle}{\abs{\nabla^\mathsf{E} \omega}}
    \end{equation}
\end{tcolorbox}

If the scalar-tensor theory is already minimally coupled so that Einstein frame amounts to $\omega = \text{const.}$, $\abs{\zeta}\geq 0$ becomes a trivial bound. Returning to our example of toroidal compactifications where $m_\text{KK}\propto e^\omega$ and independent of $\varpi$, we readily find from Eq.~\eqref{eq:sharpenedbound}
\begin{equation}
    \abs{\zeta}\geq \sqrt{\frac{1}{d-2}+\frac{1}{D-d}}\,.
\end{equation}
This bound holds for all KK towers \cite{Etheredge:2024tok}. Finally, if we include the heuristic limit $D-d\to \infty$ corresponding to string towers, we recover
\begin{equation}
    \abs{\zeta} \geq \frac{1}{\sqrt{d-2}}\,,
\end{equation}
which is none other than the bound of the SDC.

\section{Conclusion}
\label{sec:conclusion}

In this paper we have studied and expanded on the concept of frame covariance in the context of field space geometry of gravitational EFTs, with the aim of revisiting some of the Distance Conjectures within the Swampland Programme. Starting from the observation that gravitational EFTs admit multiple equivalent conformal frames related by Weyl transformations, we raised two foundational questions that are usually left implicit:
\begin{enumerate}
    \item The kinetic function $\mathsf{G}_{ij}$, conventionally taken to be the moduli space metric, transforms non-trivially under Weyl transformations relating otherwise physically equivalent conformal frames. To what extent is there a uniquely defined field space metric? Must we compute geodesics and geodesic distance in a distinguished frame? 
    \item In their current formulations, the Distance Conjectures are phrased in $d$-dimensional Planck units. How can such statements claim to encode fundamental \emph{physical} constraints related to quantum gravity while apparently depending on a choice of units? Why should $\Mpld$ play a privileged role when the physically relevant cutoff scale is often set by the species scale $\Lsp < \Mpld$?
\end{enumerate}
Clarifying these questions is central in order to understand which aspects of the Distance Conjectures are genuinely quantum-gravitational, and which instead reflect less noble, although more general, properties of gravitational EFTs.

To tackle this, we developed a new framework to systematically study the field space geometry of gravitational EFTs. The key technical step involved promoting the conformal mode to a \emph{bona fide} scalar field, thereby giving rise to a higher-dimensional frame-augmented field space $\mathcal{M}^\omega$ equipped with a single Lorentzian auxiliary metric $\mathsf{G}_{IJ}$. In this picture, Einstein frame, Jordan frame, as well as any other frame, manifest neatly as different foliations of the same higher-dimensional geometry, with their respective geometries interpreted as induced metrics $\mathsf{G}_{IJ}^\mathsf{\Sigma}$ on these hypersurfaces. This immediately resolves the first conceptual issue: there is a unique underlying metric on $\mathcal{M}^\omega$, and the apparent proliferation of frame-dependent metrics is simply a question of which slice one chooses. Among all possible foliations, we showed that the Einstein foliation is distinguished; the shift vector vanishes, the leaves are totally geodesic hypersurfaces, and hence geodesic distances in the augmented space coincide with distances computed in Einstein frame. Once the conformal factor is treated on the same footing as the other fields, the connection to units and unit transformations also becomes entirely transparent. 

We then applied this general framework --- valid for any scalar-tensor theory of the form \eqref{eq:scalartensoraction}, irrespective of UV completion --- to toroidal compactifications as a controlled example native to the Swampland Programme. Identifying the moduli space with $\mathcal{M}^\omega$, we showed that the frame-covariant definitions of the species vector $\mathcal{Z}$ and tower vector $\zeta$ reproduces known expressions. More generally, we found that the norm $\abs{\mathcal{Z}}$ is controlled by the causal character of the Jordan frame hypersurface, as summarised in Eq.~\eqref{eq:speciesscaleclaim}. In particular, the usual SSDC lower bound $\abs{\mathcal{Z}}\geq \frac{1}{\sqrt{(d-1)(d-2)}}$ is recovered only when the Jordan frame is timelike (or, when saturating the bound, null) in $\mathcal{M}^\omega$. Our suspicion is that this condition is always satisfied for theories obtained via dimensional reduction, which explains why the SSDC bound has appeared robust in a string compactification context. At the same time, our derivation demonstrates how the bound follows from general properties of scalar-tensor theories under Weyl transformations and the requirement of frame covariance, rather than from intrinsic, microscopic features of quantum gravity. 

Taken together, these results suggest a refined interpretation of the Distance Conjectures. Within our framework, one can cleanly separate the scaling of dimensionful quantities that is truly universal to gravitational EFTs from the part that is merely inherited from the chosen system of units. One can ask for example whether the SSDC upper bound ${\abs{\mathcal{Z}}\leq \frac{1}{\sqrt{d-2}}}$ or the proposed universal pattern $\zeta \cdot \mathcal{Z}=\frac{1}{d-2}$ admit similar ties to frame covariance. In addition, can it be extended to other conjectures such as the AdS Distance Conjecture? More broadly, the frame-augmented field space and its ADM-like decomposition provide a general toolkit for analysing gravitational EFTs. We hope that this perspective will help sharpen the Distance Conjectures even further, and at the same time offer a common language to disentangle the subtleties surrounding conformal frames.

\section*{Acknowledgements}
We would like to thank Bruno Bento, Martin Carrascal, Zongzhe Du, Antonio Padilla, Susha Parameswaran, Kajal Singh, and Kieran Wood for many useful discussions and comments on the draft. BM would also like to thank Osmin Lacombe, Shinji Mukohyama, and Josef Seitz for their collaboration during the early stages of this work. SK was supported by the Estonian Research Council Mobilitas 3.0 incoming postdoctoral grant MOB3JD1233 ``Inflationary Nonminimal Models: An Investigative Exploration''.

\appendix
\section*{Appendix}
\section{Examples}\label{sec:appendix}
In this brief appendix we show how the result in Eq.~\eqref{eq:speciesscaleclaim} applies to some scalar-tensor models. To match the convention used in the main part of the paper we pull out a dimensionful factor to make the scalar fields dimensionless.

\paragraph{Brans-Dicke Theory.} The Brans-Dicke action in $d$ dimensions reads
\begin{equation}
    S_\text{BD} = \int\dd^dx \sqrt{-g}\ \mathrm{M}^{d-2}\left[\frac{1}{2}\phi R - \frac{1}{2}\frac{\omega_\text{BD}}{\phi}(\partial\phi)^2\right]\,.
\end{equation}
From this we can read off $f(\phi) = \mathrm{M}^{d-2}\phi$ and thus $M_\mathsf{J} = \mathrm{M}\phi^\frac{1}{d-2}$. Working in Planck units $M_\mathsf{E}=\mathrm{M}$, the dimensionless field space metric in Einstein frame is
\begin{equation}
    \mathsf{G}_{\phi\phi}^\mathsf{E} = \frac{\omega_\text{BD}}{\phi^2} + \frac{d-1}{d-2}\frac{1}{\phi^2}\,.
\end{equation}
This places a bound $\omega_\text{BD} > -\frac{d-1}{d-2}$ in order for the Einstein frame metric to be Euclidean. One straightforwardly computes the analogous species vector
\begin{equation}
    \mathcal{Z}_\phi = \frac{1}{M_\mathsf{J}}\mathsf{D}_\phi^\mathsf{E}M_\mathsf{J} = \frac{1}{d-2}\frac{1}{\phi}
\end{equation}
and its norm
\begin{equation}
    \abs{\mathcal{Z}} = \frac{1}{d-2}\sqrt{\frac{1}{\omega_\text{BD}+\frac{d-1}{d-2}}} > 0\,.
\end{equation}
We can distinguish three cases:
\begin{itemize}
    \item $\omega_\text{BD}>0:$ one finds that Jordan frame is spacelike and $\abs{\mathcal{Z}} < \frac{1}{\sqrt{(d-1)(d-2)}}$.
    \item $\omega_\text{BD}=0:$ the induced Jordan frame metric vanishes $\mathsf{G}_{\phi\phi}^\mathsf{J}=0$ as the Jordan frame becomes a null hypersurface. The species vector norm is exactly $\abs{\mathcal{Z}}=\frac{1}{\sqrt{(d-1)(d-2)}}$.
    \item $0>\omega_\text{BD}>-\frac{d-1}{d-2}:$ Jordan frame becomes timelike and $\abs{\mathcal{Z}}> \frac{1}{\sqrt{(d-1)(d-2)}}$.
\end{itemize}
All is in agreement with Eq.~\eqref{eq:speciesscaleclaim}.

\paragraph{Higgs Inflation.} Next we look to the singlet Higgs Inflation model generalised to $d$ dimensions
\begin{equation}
    S_\text{HI} = \int \dd^d x \sqrt{-g}\ \mathrm{M}^{d-2}\left[\frac{1}{2}(1+\xi \abs{H}^2)R - \frac{1}{2}\abs{\partial H}^2\right]
\end{equation}
with $H = he^{i\phi}$. In this case, the Jordan frame is always spacelike for $\xi>0$. The only non-vanishing component of the species vector is
\begin{equation}
    \mathcal{Z}_h = \frac{1}{d-2}\frac{2\xi h}{1+\xi h^2}
\end{equation}
and its norm is
\begin{equation}
    \abs{\mathcal{Z}} = \frac{1}{d-2}\sqrt{\frac{(2\xi h)^2}{1+\xi h^2} \frac{1}{1+\frac{d-1}{d-2}\frac{(2\xi h)^2}{1+\xi h^2}}} < \frac{1}{\sqrt{(d-1)(d-2)}}
\end{equation}
in agreement with the claim.

\bibliographystyle{JHEP}
\bibliography{bibliography}

\end{document}